\documentclass[aps,prl,superscriptaddress,twocolumn,showpacs,longbibliography]{revtex4-1}
\usepackage{amsmath}
\usepackage{amsthm, amssymb} 
\usepackage{bbold}                       
\usepackage{graphicx}
\usepackage{color}
\usepackage{times, verbatim}      

\usepackage{hyperref}
\hypersetup{colorlinks=true}

\newcommand{\BEq}{\begin{eqnarray}}
\newcommand{\EEq}{\end{eqnarray}}

\begin{document}

\title{Particle-hole-symmetric model for a paired fractional quantum Hall State in a half-filled Landau level}

\affiliation{Department of Physics \& Astronomy, California State University Long Beach,  Long Beach, California 90840, USA}
\affiliation{Department of Physics, Virginia Tech, Blacksburg, Virginia 24061, USA}
\affiliation{Shenzhen Institute for Quantum Science and Engineering, and Department of Physics, Southern University of Science and Technology, Shenzhen 518055, China}
\affiliation{Kavli Institute for Theoretical Physics, University of California, Santa Barbara, California 93106, USA}

\author{William Hutzel$^1$}
\author{John J. McCord$^1$}
\author{P.T. Raum$^2$}
\author{Ben Stern$^2$}
\author{Hao Wang$^3$}
\author{V. W. Scarola$^2$}
\author{Michael R. Peterson$^{1,4}$}

\begin{abstract}
The fractional quantum Hall effect (FQHE) observed at half filling of the second Landau level is believed to be caused by a pairing of composite fermions captured by the Moore-Read Pfaffian wave function.  The  generating Hamiltonian for the Moore-Read Pfaffian is a purely three-body model that breaks particle-hole symmetry and  lacks other properties, such as dominate two-body repulsive interactions, expected from a physical model of the FQHE. We use exact diagonalization to study the low energy states of a more physical two-body generator model derived from the three-body model.  We find that the two-body model exhibits the essential features expected from the Moore-Read Pfaffian: pairing, non-Abelian anyon excitations, and a neutral fermion mode.  The model also satisfies constraints expected for a physical model of the FQHE at half-filling because it is: short range, spatially decaying, particle-hole symmetric, and supports a roton mode with a robust spectral gap in the thermodynamic limit. Hence, this  two-body model offers a bridge between artificial three-body generator models for paired states and the physical Coulomb interaction and can be used to further explore properties of non-Abelian physics in the FQHE. 
\end{abstract}

\date{November 6, 2018}


\maketitle

\section{Introduction}	

Soon after the construction of the Laughlin wave functions\cite{Laughlin83} for the FQHE~\cite{Tsui82} at electronic filling factors $\nu=1/(2p+1)$ ($p$ an integer), a short-range generator Hamiltonian \cite{Haldane83,Trugman85} was found that produced the Laughlin wave functions as unique gapped ground states.  This  model shared properties of the Coulomb interaction: It is two-body, consists of interactions  decaying with distance, and is invariant under particle-hole (PH) transformations.  Moreover, the model was  shown to generate states that accurately described the experimentally observed  FQHE at filling factors $\nu=n_0/(2pn_0\pm1)$ (for integer $n_0$) in the lowest Landau level (LLL).  In fact, the ground states of this generator model are virtually identical to the composite fermion (CF) wave functions \cite{Jain89,jain2007composite} (the CF wave functions incorporate the Laughlin wave functions as a subset).  The CF wave functions are written as $\mathcal{J}^{2p}\phi$, where the Jastrow factor $\mathcal{J}$ ``binds" $2p$ vortices of the many-body wave function to electrons described by $\phi$.  This strongly interacting \emph{electron} wave function $\mathcal{J}^{2p}\phi$ is interpreted as a wave function for  CFs described only by $\phi$.  The choice $\phi \rightarrow \phi_{n_0}$ describes CFs completely filling $n_0$ CF LLs and yields low-energy wave functions with the same quantum numbers and physics of the low-energy states  of the generator model, the Coulomb interaction in the lowest LL, and  importantly describes the FQHE at  filling factor $\nu=n_0/(2pn_0\pm1)$.  These wave functions (and generator model) also predict a gapless 
PH-symmetric state at half-filling described as a CF-Fermi sea ($\phi \rightarrow \phi_{\text{FS}}$)~\cite{Halperin93,Rezayi94,Rezayi00,Geraedts16}
that accurately captures the  physics of the Coulomb interaction at half-filling of the LLL~\cite{Willett93,jain2007composite}.

The unexpected discovery of the FQHE in the half-filled second LL \cite{Willett87} (total filling factor $5/2$) led to the construction of the gapped  Moore-Read Pfaffian state at half-filling \cite{Moore91}.  This state can be interpreted as a paired state of CFs~\cite{Scarola00,Moller2008}, written via  $\phi \rightarrow\phi_{\text{BCS}}$ where $\phi_{\text{BCS}}$ is a Bardeen-Cooper-Schrieffer  state of CFs (a Pfaffian in real space) pairing CFs in the $p$-wave channel~\cite{Moore91,Read00}. Wave functions of this type are excellent candidates for the FQHE at filling factor $5/2$ \cite{Morf98,Rezayi00,Peterson08,Peterson08b,Feiguin08,Feiguin09,Wojs10,Storni10,Rezayi2011,Pakrouski2015,Zaletel15,Tylan2015,Rezayi2017} and predict non-Abelian quasiquasiparticlesparticles \cite{Moore91,Nayak96c,Tserkovnyak03,Baraban09,Bonderson11a}, which, if identified experimentally, could form building blocks in the construction of a topologically protected quantum computer~\cite{DasSarma05,Nayak08}. 

Interestingly, a purely repulsive three-body generator Hamiltonian \cite{Greiter91} (labelled $H_3$) yields most of the physics described by Moore and Read \cite{Moore91}.  Specifically, $H_3$ generates the Moore-Read Pfaffian wave function as an exact ground state and produces a degenerate manifold of non-Abelian quasi-hole excitations  \cite{Read96}.  But $H_3$ does not obey \emph{all}  the properties expected of a \textit{physical}  model of the FQHE at  half-filling.  First, the model does not respect PH-symmetry.  This constraint might not be crucial since numerical work indicates the ground state of the Coulomb interaction in the half-filled second LL breaks PH-symmetry \cite{Wang2009} with additional PH-symmetry breaking terms. Importantly, LL-mixing effects in realistic models   supply emergent PH-symmetry breaking three-body terms \cite{Bishara09a,Peterson13b,Sodemann2013,Rezayi2013,Wooten13}.  Second, $H_3$ is purely three-body, challenging theory to bridge from it to physical two-body dominate models since additional two-body terms added to $H_3$ generally lift \cite{Toke06a,Peterson08,Wang2009,Storni10} expected degeneracies \cite{Read96}. The construction of $H_3$ is constrained by the Pfaffian form of $\phi_{\text{BCS}}$ but other forms\cite{Read00}  could yield states and generator models  satisfying all physical requirements while potentially preserving  topological properties including non-Abelian anyon excitations.

The authors of Ref.~[\onlinecite{Peterson08c}] discovered that $H_3$  can be added to its PH conjugate Hamiltonian $\overline{H}_3$ (which generates the  anti-Pfaffian~\cite{Lee07,Levin07}) to yield a purely $\emph{two-body}$ model 
\BEq
H_2 = H_3 + \overline{H}_3\;.
\label{eq_H2}
\EEq
$H_2$  has all of the properties desired of a physical model: It is two-body and PH-symmetric and spatially decays with distance. Furthermore, the ground state energy of $H_2$, as a function of particle number, displays a ``wine bottle" potential structure, interpreted originally as evidence for spontaneous PH-symmetry breaking in the ground state.  Subsequent work  showed the ground state energies have a prominent even-odd effect \cite{Lu2010} indicative of pairing.  These properties are consistent with the model's definition in terms of generators for the Moore-Read Pfaffian/anti-Pfaffian.  However, many important aspects of $H_2$ remain unexplored as a stand-alone generator model for FQHE at half-filling of a single LL.  

In this work we use numerical exact diagonalization to show that $H_2$ offers a more physical generator model for the FQHE  at half-filling that shares  essential features of the generator model of the Moore-Read Pfaffian including a spectral gap at half-filling with a neutral roton mode \cite{Moller2011,Wurstbauer2013}.  When combined with the fact that $H_2$ is short ranged, PH-symmetric, and two-body, it becomes a useful generator model for half-filled FQHE states that connects the physical Coulomb interaction with the non-physical three-body generator model of the Moore-Read Pfaffian.  Furthermore, we show  the low-energy states of $H_2$ are adiabatically connected to the Moore-Read Pfaffian, possess the same topological entanglement properties, and support (quasi)-degenerate non-Abelian quasiparticles.  We find conflicting evidence for spontaneous PH-symmetry breaking in the ground state of $H_2$.  More work is needed to unambiguously determine if the ground state of $H_2$ spontaneously break PH-symmetry at half filling.  

The paper is organized as follows.  In Sec.~\ref{sec_model} we define the two-body model, $H_2$, and show how it can be rewritten as a linear combination of three-body models.  Section~\ref{sec_excitations} examines the low-energy excitations and shows that $H_2$ supports a FQHE gap at half-filling in the thermodynamic limit.  The roton mode and neutral fermion mode, expected from the Moore-Read Pfaffian, are shown to exist.  Section~\ref{sec_adiabatic} tracks the low-energy excitations while  tuning between $H_2$ and $H_3$.  Here it is shown that all low-energy states are adiabatically connected and  the low-energy manifold of $H_2$ possess (quasi)-degenerate non-Abelian quasi-holes consistent with Moore-Read Pfaffian expectations.  Section~\ref{sec_phs} studies the PH-symmetry properties of the ground state of $H_2$.  It is found that the ground states using the torus geometry do not appear to spontaneously break PH-symmetry, but further work is needed to conclusively establish this fact. Finally, Sec.~\ref{sec_entanglement} shows  the ground and low-energy states have entanglement properties consistent with the Moore-Read Pfaffian.  We summarize our results in Sec.~\ref{sec_summary}.    

\section{Model}	
\label{sec_model}

The two-body model we consider is a short-range model of $N$ interacting  spin-polarized fermions \cite{footnote} confined to two-dimensions and the LLL:
\begin{eqnarray}
H_2 &=& \sum_{i<j}^N\left[ \hat{P}_{ij}(1)+\frac{1}{3}\hat{P}_{ij}(3)\right] \nonumber\\
	&=& \sum_{i<j<k}^N\left[\hat{P}_{ijk}(3)+\overline{\hat{P}}_{ijk}(3)\right],
    \label{eq_H_map}
\end{eqnarray}
where  $\hat{P}_{ij}(m)$ denotes projection \cite{Haldane83} onto two-body eigenstates of relative angular momentum $m$.   Similarly, $\hat{P}_{ijk}(m)$ denotes projection onto the \emph{three-body} eigenstates of relative angular momentum.  Here and in what follows, the overline denotes PH conjugation and we focus on half-filling.  We work in energy units of the interaction strength and all distances are in units of the magnetic length.

The first line in Eq.~(\ref{eq_H_map}) shows that $H_2$ is a repulsive two-body interaction that decays with inter-particle separation.  We can see this by noting that inter-particle separation increases with $m$.  More explicitly, the two-body projectors can be written in real-space in the disk geometry\cite{Trugman85} as
\begin{eqnarray}
\hat{P}_{ij}(m)= \nabla^{2m}\delta(r_{ij})
\end{eqnarray}
where $r_{ij}$ denotes the planar separation between particles $i$ and $j$.  The $m=3$ term in Eq.~(\ref{eq_H_map}) enforces repulsion at distances larger than the $m=1$ term alone.  At half-filling the $m=1$ term, by itself, is known to generate the CF Fermi sea \cite{Halperin93,Rezayi94,jain2007composite}.  We show below  the addition of the second $m=3$ term leads to pairing.  The prefactor of $1/3$ on the $m=3$ term derives automatically from a re-expression of the two-body interaction as a sum of three-body interactions \cite{Peterson08c}.   Appendix~\ref{sec_haldane} discusses the Haldane pseudopotential expansion of $H_2$ for finite-sized spherical systems \cite{Haldane83}.

The second line in  Eq.~(\ref{eq_H_map}) shows the remarkable fact that a repulsive two-body interaction can be rewritten as the exact generator of CF paired states.  $H_3\equiv \sum_{i<j<k}\hat{P}_{ijk}(3)$ and $\overline{H}_3\equiv \sum_{i<j<k}\overline{\hat{P}}_{ijk}(3)$ are Hamiltonians that generate the Moore-Read Pfaffian $\Psi_\mathrm{Pf}$ and its PH conjugate, the anti-Pfaffian  $\Psi_\mathrm{aPf}\equiv\overline{\Psi}_\mathrm{Pf}$, respectively [more compactly written in Eq.~(\ref{eq_H2})].  The equality in Eqs.~(\ref{eq_H2}) and (\ref{eq_H_map}) hold up to single particle terms that we have absorbed into a redefined chemical potential.  

An important feature of $H_2$ is that it precisely connects the CF-Fermi sea  [specifically ground states of $\sum_{i<j} \hat{P}_{ij}(1)$] with the Moore-Read Pfaffian.  Over-screening of the inter-CF interaction can lead to a Kohn-Luttinger-type \cite{Kohn65,Chubukov93} instability in the CF-Fermi sea toward a CF paired state thereby favoring wave functions with $\phi_{\text{BCS}}$ over $\phi_{\text{FS}}$. The CF wave functions themselves were shown to harbor their own instability in the $p$-wave channel when studied in the second LL \cite{Scarola00}. In other words, $H_2$ shows that by adding $\hat{P}_{ij}(3)$ terms to the model that generates the CF-Fermi sea, a paired state is favored.  At lowest-order, when electrons form CFs, the vortex binding accommodates the energy cost of the $\hat{P}_{ij}(1)$ term at short range. The addition of $\hat{P}_{ij}(3)$ terms can be interpreted as forcing an over-screening of the $\hat{P}_{ij}(1)$ interaction terms leading to a pairing instability of the CF Fermi sea. 

\section{Low-energy excitations and the energy gap}
\label{sec_excitations}

\begin{figure}[ht!]
\begin{center}
\includegraphics[width=8cm,angle=0]{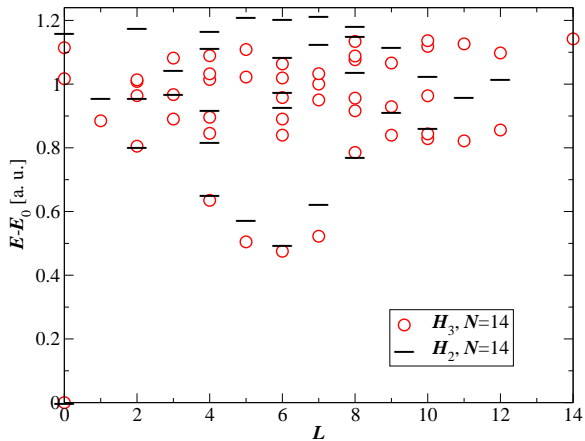}
\caption{(Color online)  Energy spectrum of the two-body model, $H_2$ (black dashes), and three-body model that generates the Moore-Read Pfaffian, $H_3$ (red circles), for $N=14$ and $Q=12.5$ on the sphere.  }
\label{fig:spectrum}
\end{center}
\end{figure}

\begin{figure}[ht!]
\begin{center}
\includegraphics[width=8cm,angle=0]{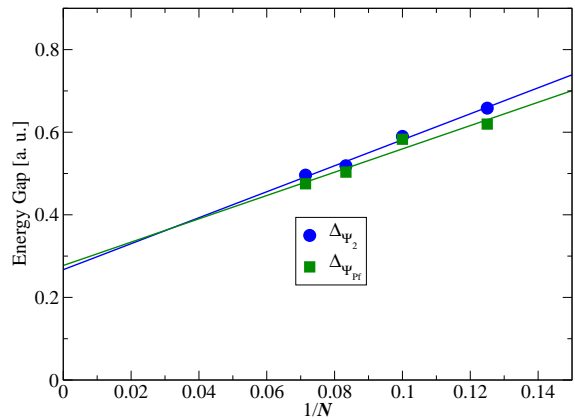}
\caption{(Color online)  Energy gap between the lowest energy excited state and the uniform ground state of $H_2$ (denoted by $\Delta_{\Psi_2}$) and $H_3$ (denoted $\Delta_{\Psi_{\text{Pf}}}$), respectively, as a function of inverse particle number.  Linear extrapolations find the gaps in the thermodynamic limit to be $\Delta_{\Psi_2}=0.267(24)$ and $\Delta_{\Psi_\mathrm{Pf}}=0.277(47)$.  The numbers in parenthesis are the standard deviation in the linear extrapolation.}
\label{fig:gap}
\end{center}
\end{figure}

We begin our study of $H_2$ by addressing the FQHE energy gap  using exact diagonalization in the spherical geometry \cite{Haldane83}. Half-filling occurs for $N=(2Q + S)/2$ where $2Q$ is the total magnetic flux through the surface of a sphere of radius $R=\sqrt{Q}$ and $S$ is the so-called shift, an order-one correction that vanishes in the thermodynamic limit \cite{Wen90b}.  The ground state of $H_3$ ($\overline{H}_3$) is  $\Psi_\mathrm{Pf}$ ($\Psi_\mathrm{aPf}$) at a shift of $S=3$ ($S=-1$). A gap is necessary for the ground state of $H_2$ at $2Q=2N-S$ to represent a valid FQH state.  Hence, we calculate the low-energy spectrum of $H_2$ (shown  for $N=14$ in Fig.~\ref{fig:spectrum} and in Ref.~[\onlinecite{Peterson08c}] for $N=8$) and define the gap as the difference between the energy of the first excited state and the $L=0$ ground state (if the ground state has $L\neq 0$ then the gap is take to be zero).  $\Psi_2$ is found to be a uniform state with total angular momentum $L=0$ separated from  excited states by a finite gap, $\Delta_{\Psi_2}$.  In fact, the structure of the low-energy spectrum is notably similar to the low-energy spectrum of the second LL Coulomb interaction and $H_3$~\cite{Peterson08c}.  We also calculate the thermodynamic limit of the energy gap between the first excited state and the $L=0$ ground state--the so-called ``roton" gap.  From Fig.~\ref{fig:gap} we see  the gap is  finite and nearly identical in the thermodynamic limit to the Moore-Read Pfaffian gap, $\Delta_{\Psi_\mathrm{Pf}}$ \cite{Storni10}.

\begin{figure}[ht!]
\begin{center}
\includegraphics[width=8cm,angle=0]{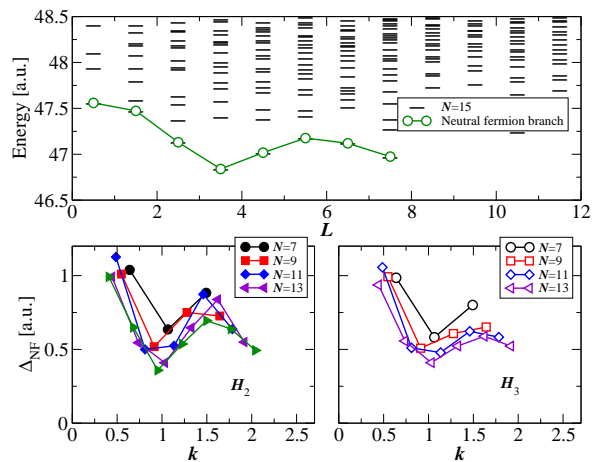}
\caption{(Color online) The top panel shows the spectrum of $H_2$ for $N=15$ at $2Q=27$ and identifies the neutral fermion mode.  The bottom panels show the neutral fermion modes for systems up to $N=13$ for $H_2$ (left) and  $H_3$ (right).  These figures can be compared to those for $H_3$ and the second LL Coulomb Hamiltonian in Ref.~[\onlinecite{Moller2011}].  The wave vector is $k=L/\sqrt{Q}$. 
}
\label{fig-NF}
\end{center}
\end{figure}

The Moore-Read Pfaffian state additionally supports a so-called neutral fermion mode~\cite{Moore91,Read00,Bonderson11c,Moller2011} which we can study in $H_2$ as well.  Following Ref.~[\onlinecite{Moller2011}] we calculate the neutral fermion mode by considering a system at odd $N$ and $2Q=2N-3$.  To define the neutral gap we construct a ``ground state'' energy at odd $N$ by finding the linear interpolation between the ground state energy of the nearby even particle systems at $N+1$ and $N-1$ (again for $2Q=2N-3$).  The neutral mode dispersion $\Delta_\mathrm{NF}(k)$ of $H_2$ is shown in Fig.~\ref{fig-NF} and is remarkably similar, qualitatively and quantitatively, to the neutral mode of $H_3$.  

\section{Adiabatic continuity}
\label{sec_adiabatic}

To investigate whether $\Psi_2$ is indeed in the same universality class as the Moore-Read Pfaffian we  consider the adiabatic continuity (or lack thereof) between the ground and low-energy states of $H_2$ to those of $H_3$.  More concretely, we consider the Hamiltonian
\BEq
H(\alpha) = (1-\alpha)H_3 + \alpha H_2
\EEq
that interpolates between $H_3$ and $H_2$ for $\alpha\in[0,1]$.  Clearly, $H(0) = H_3$ and $H(1)=H_2$ where the ground states are $\Psi_\mathrm{Pf}$ and $\Psi_2$, respectively.  As we tune $\alpha$ from zero to unity we  track the ground state and energy gap. We denote the ground state(s) of $H(\alpha)$ as $\Psi$.

\subsection{Adiabatic continuity in the ground state and neutral modes}

We first consider the spherical geometry at $2Q = 2N- 3$ and investigate the ground states of $H(\alpha)$.  For the two systems  to be adiabatically connected  we expect the spectrum to maintain a uniform ground state with $L=0$ in addition to a finite gap $\Delta$ that smoothly interpolates between the two end points without vanishing.  Otherwise we expect the gap to close at some finite $\alpha_c$ indicating a quantum phase transition between $\Psi_\mathrm{Pf}$ and $\Psi_2$. 

\begin{figure}[ht!]
\begin{center}
\includegraphics[width=8cm,angle=0]{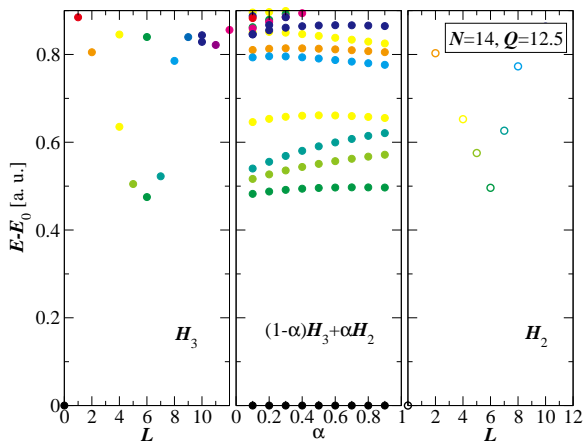}
\caption{(Color online)  Energy of $H(\alpha)$ relative to the ground state $E_0$.  The middle panel shows the low-energy spectrum (lowest approximately 15 states) for $H(\alpha)$ as a function of $\alpha$ for $N=14$ and $Q=12.5$.  The left and right panels plot relative energy as a function of angular momentum $L$  for $\alpha=0$ and 1, respectively.  The angular momentum of each state is indicated by color.  The gap stays open and relatively constant indicating adiabatic continuity.
}
\label{fig-ac-gs}
\end{center}
\end{figure}

Figure~\ref{fig-ac-gs} shows the low-energy spectrum for $H(\alpha)$ as a function of $\alpha$ for $N=14$.  The low-energy states of $H_3$ and $H_2$ are clearly adiabatically connected--the gap $\Delta$ between the first excited state and the ground state remains open and remarkably constant from $\alpha=0$ to $\alpha=1$.  The absolute size of $\Delta$ for $H_3$ and $H_2$ are within $5\%$ of one another at the end points.  Furthermore, many of the higher-energy states (in the continuum)  are also adiabatically connected.  Smaller systems ($N=12$, 10, and 8) show similar qualitative and quantitative results.  

\begin{figure}[ht!]
\begin{center}
\includegraphics[width=7.5cm,angle=0]{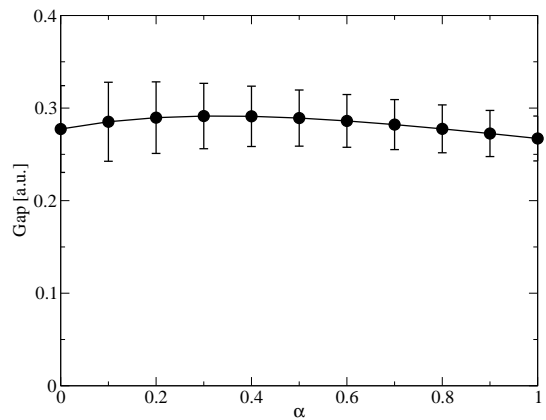}
\caption{(Color online)  The thermodynamic limit of the energy gap of $H(\alpha)$ is shown versus $\alpha$.  Similar to the finite-size system results of Fig.~\ref{fig-ac-gs} the gap remains finite and largely flat, adiabatically connecting $\Psi_2$ with $\Psi_\mathrm{Pf}$.  The error-bars indicate the standard deviation in the linear extrapolation.  
}
\label{fig-ac-gs-therm}
\end{center}
\end{figure}

Figure~\ref{fig-ac-gs} represents adiabatic continuity between $\Psi_2$ and $\Psi_\mathrm{Pf}$ for finite sized systems (we have shown $N=14$). However,  to examine the effect of system size on the adiabatic continuity we calculate the linear extrapolation of the energy gap versus $1/N$, i.e., we take the thermodynamic limit, for several $\alpha$ between zero and one.   Figure~\ref{fig-ac-gs-therm} shows  the thermodynamic limit of $\Delta$ remains finite for all $\alpha$ and essentially reflects the finite-sized system results.   The fact the system remains gapped in this limit further supports the conclusion that $\Psi_2$ and $\Psi_\mathrm{Pf}$ are in the same universality class.

\begin{figure}[ht!]
\begin{center}
\includegraphics[width=8cm,angle=0]{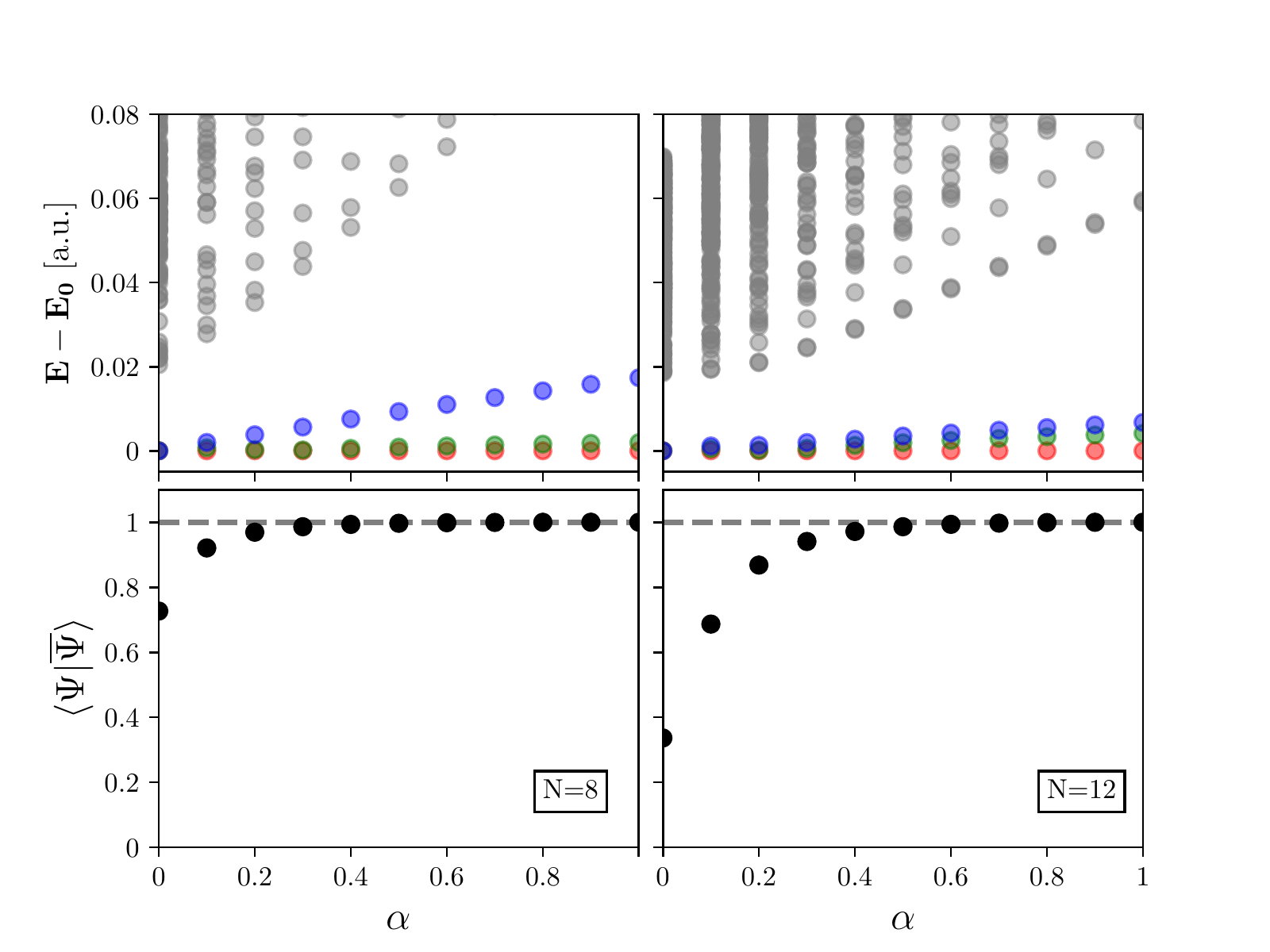}
\caption{(Color online) Relative energy of $H(\alpha)$ in the torus geometry for $N=8$ (upper-left panel) and $N=12$ (upper-right panel) for an aspect ratio $\tau=0.95$ as a function of $\alpha$. The threefold ground-state degeneracy remains quasidegenerate, and well below the gap, as $\alpha$ is tuned from $H_3$ to $H_2$.  The bottom panels show the corresponding overlap $\langle \Psi | \overline{\Psi}\rangle$ indicating that $\Psi_2$ is PH-symmetric at $\alpha=1$ since $\langle \Psi_2 |\overline{\Psi_2}\rangle=1$, $\Psi_\mathrm{Pf}$ breaks PH-symmetry since $\langle \Psi_\mathrm{Pf}|\overline{\Psi_\mathrm{Pf}}\rangle\neq 1$, and $\Psi$, the ground state of $H(\alpha)$, remains largely PH-symmetric for finite $\alpha$ less than unity.
}
\label{fig-PHSTorus}
\end{center}
\end{figure}

We now use exact diagonalization on the torus to further investigate the low energy states of $H(\alpha)$.  We work with the rectangular unit cell with aspect ratio $\tau$ near unity and present results for system sizes with ground states at the total momenta consistent with pairing, i.e., the same ground state momenta found for the ground states of $H_3$ (cf. Refs.~\onlinecite{Greiter92,Peterson08}).  The upper two panels of Fig.~\ref{fig-PHSTorus} show the low-energy spectrum of $H(\alpha)$  for $N=8$ and $N=12$ with  $\tau=0.95$ (our results are robust to changes in $\tau$).  In this geometry the topological order of the Moore-Read Pfaffian state is in evidence by the existence of a three-fold ground-state degeneracy separated from  the higher-energy continuum by a gap.  As $\alpha$ is tuned from $H_3$ to $H_2$ we see that, while the three-fold ground-state degeneracy is minimally broken due to ``tunneling" between topological sectors, the threefold quasidegeneracy remains well-below the continuum states all the way to $H_2$. These results are qualitatively similar to those using the spherical geometry (Fig.~\ref{fig-ac-gs}) and lend even more support for the adiabatic continuity between $\Psi_2$ and $\Psi_\mathrm{Pf}$.  For $N=10$ and $N=14$, we did not find a paired ground state on the torus for $H_2$ with the rectangular unit cell.  For these particle numbers the ground state is not threefold degenerate and occurred at wave vectors different from the paired states.  This could be due to a finite size effect which favors non-uniform states on the torus for the $H_2$ model with the rectangular unit cell \cite{Rezayi00}.  

\subsection{Adiabatic continuity in the quasi-hole sector: Non-Abelian anyons in a particle-hole-symmetric model}

The quasihole sector of $H_3$ supports non-Abelian excitations that can be utilized as building blocks for a topological quantum computer \cite{Nayak08}.  To search for the same feature in $H_2$ we study the quasiholes in the spherical geometry where $2Q=2N-2$ is a system with two non-Abelian quasi-hole excitations.  In this geometry, the non-Abelian nature of the excitations manifests through the existence of a zero-energy manifold of states \cite{Read96}.  For topological quantum computing applications it is important that the non-Abelian excitations be degenerate, or at least, the quasi-degenerate states must be significantly below the energy gap to generic excitations such that at low temperatures and weak disorder there is an exponentially suppressed probability of the system exciting  generic excitations. 

\begin{figure}[ht!]
\begin{center}
\includegraphics[width=8cm,angle=0]{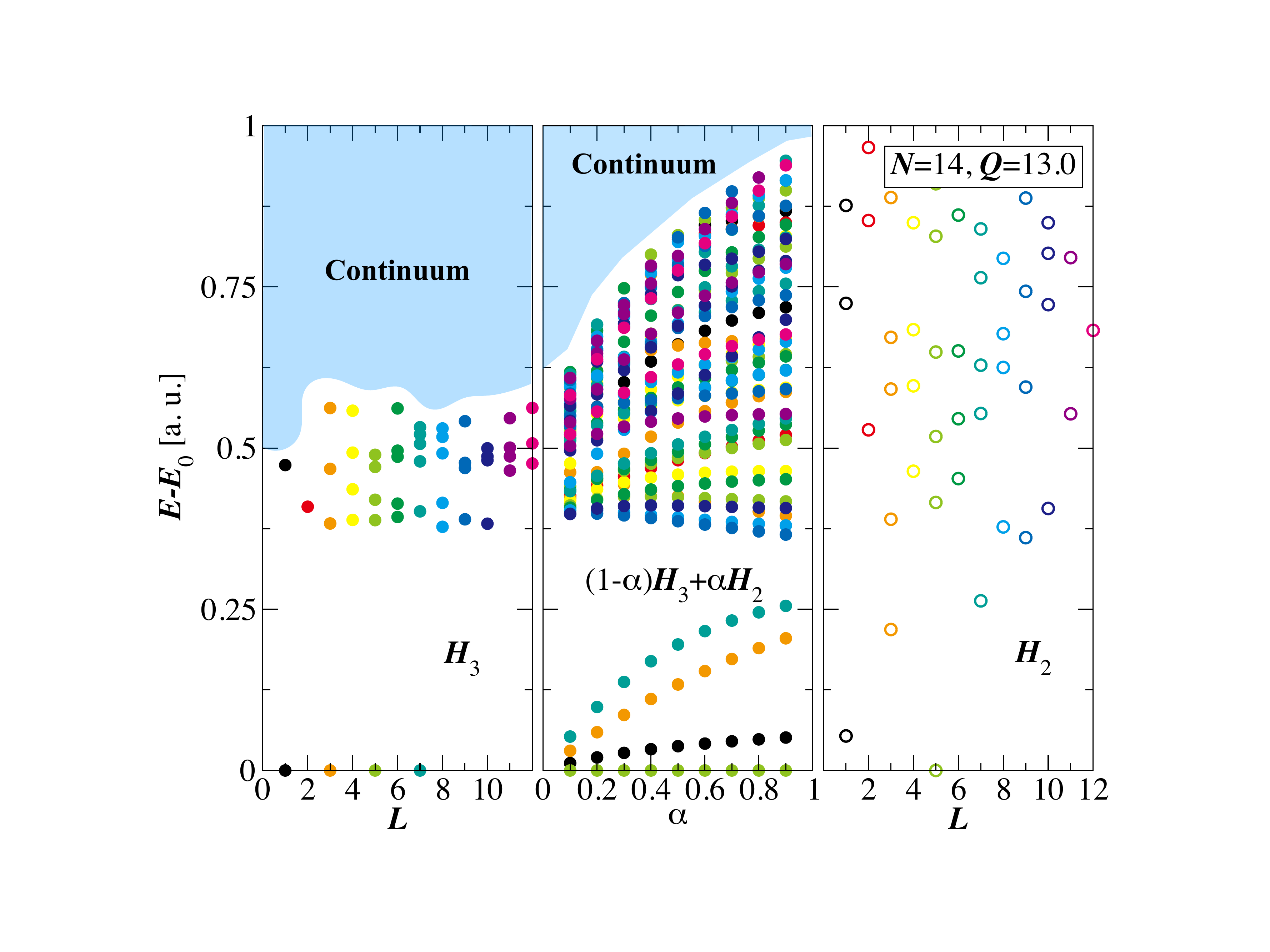}
\caption{(Color online) Energy of $H(\alpha)$ relative to the ground state (see Fig.~\ref{fig-ac-gs}) for the system containing two quasi-hole excitations ($N=14$ and $Q=13$) (the blue sections on the left and middle panels indicate there are higher energy states in the continuum we did not calculate).  Nonzero $\alpha$ causes the zero-energy degenerate non-Abelian quasihole manifold to be broken; however, the spread of states stays below the continuum of generic excitations all the way to $\alpha=1$. 
}
\label{fig-ac-qh}
\end{center}
\end{figure}

In Fig.~\ref{fig-ac-qh} for $\alpha=0$ the degenerate manifold of zero-energy states of $H_3$ is clearly visible.  When 
 $\alpha\neq 0$ the degeneracy of the zero-energy  manifold is broken by adding any amount of $H_2$.  However, even as $\alpha\rightarrow 1$ the spread of the quasidegenerate manifold stays well below the gap to generic excitations.  Hence, even the exactly degenerate non-Abelian quasihole states of $H_3$ are adiabatically connected to the quasidegenerate non-Abelian quasihole states of $H_2$ for finite sized systems ($N=8$, 10, 12 show similar behavior).   Again, as we did when studying the ground state sector, we  investigate the thermodynamic limit of this apparent adiabatic continuity of the two quasihole sector.  

While it is true that the quasidegenerate manifold of quasi-hole states remains below the continuum for all system sizes investigated and adiabatic continuity appears manifest, we do observe the degeneracy of the quasihole states of $H_3$  to be  broken upon the inclusion of the two-body term of $H_2$.  Also, the spread in energy of the quasidegenerate states monotonically increases with $\alpha$. To investigate this in more detail we define $\delta$ to be the average energy of the quasidegenerate quasihole manifold of $H(\alpha)$ and $\Delta$ to be the gap between the lowest energy state in the continuum and $\delta$.

\begin{figure}[ht!]
\begin{center}
\includegraphics[width=7.5cm,angle=0]{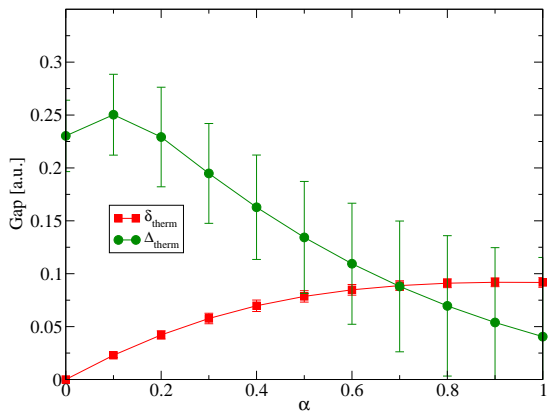}
\caption{(Color online)  The thermodynamic limit of the quasi-hole average energy,   $\delta_\mathrm{therm}$ (red), and the energy gap to the generic continuum of states, $\Delta_\mathrm{therm}$ (green), as a function of $\alpha$. At $\alpha\sim 0.7$ the spread of the quasi-degenerate quasi-holes bleeds into the continuum and appears to indicate the adiabatic continuity between the low energy states in the quasi-hole sector of $H_2$ and $H_3$ is lost.  
}
\label{fig-will-ratios}
\end{center}
\end{figure}

As we tune $\alpha$ close to $H_2$ (Fig.~\ref{fig-will-ratios}), the thermodynamic limit of $\delta$ ($\delta_\mathrm{therm}$) saturates to a value well below the thermodynamic limit of the neutral gap.  But $\Delta_\mathrm{therm}$ (the thermodynamic limit of $\Delta$), by contrast, is  reduced with increasing $\alpha$. Moreover,  $\delta_\mathrm{therm}$ becomes larger than $\Delta_\mathrm{therm}$ for $\alpha \gtrsim 0.6$--$0.8$ indicating   for infinite system sizes the gap between the ground state and low lying excited states closes for two-body interactions.  Naively this would mean the low-energy states of $H_3$ and $H_2$ for two quasiholes are not adiabatically connected. It appears the  loss of adiabatic continuity between $H_3$ and $H_2$ in the quasihole sector is at odds with the ground state sector which shows clear and robust adiabatic continuity. Or, perhaps,  purely two-body interactions cannot host non-Abelian quasihole excitations. But, as we  argue below, the transition to  $\delta_\mathrm{therm}/\Delta_\mathrm{therm}>1$ is due to a finite-size effect limitation of our exact calculations. 

The breaking of the degeneracy of the non-Abelian quasiholes can be impacted by at least two causes.  First,  the low energy states are simply not adiabatically connected. However, Fig. \ref{fig-ac-qh}, while showing a spreading of the quasidegenerate states with $\alpha$, clearly shows the same states staying below the continuum.  Appendix \ref{sec_tracking} also shows a detailed tracking of individual states and, again, the states are adiabatically connected and never mix or cross higher energy generic states in the continuum.  The quasidegenerate states of $H(\alpha)$ also maintain high overlaps with the exactly degenerate quasihole states of $H_3$ (see Appendix \ref{sec_tracking}).  

\begin{figure}[ht!]
\begin{center}
\includegraphics[width=7.5cm,angle=0]{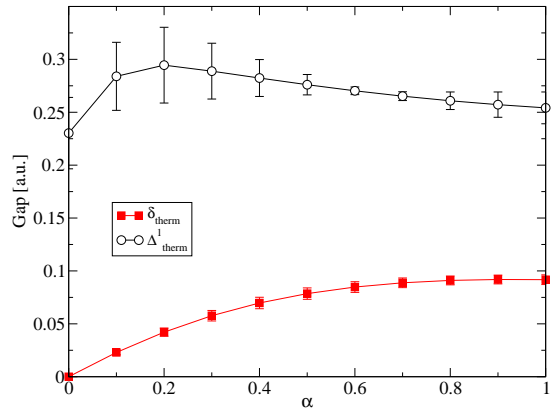}
\caption{(Color online)  The thermodynamic limit of the gaps between the lowest energy state in the continuum and the smallest angular momentum states of the quasidegenerate manifold $\Delta^1_\mathrm{therm}$, i.e., $L = 0$, 1, 0, and 1 for $N = 8$, 10, 12, and 14, respectively. This gap (black open circles) is well behaved and linear in $1/N$ as indicated by the small error bars in the extrapolation.   Finite-size effects have been greatly reduced in comparison to Fig.~\ref{fig-will-ratios} and the spread in the (quasi)degenerate quasihole energy remains below the gap to generic states in the continuum. 
}
\label{fig-qh-therm}
\end{center}
\end{figure}

The second possible cause of the broken degeneracy in the quasihole manifold stems from a finite-size effect. The quasiholes themselves are finite and can overlap in real-space.  The addition of the two-body interaction term in $H_2$ produces an energy cost for this overlap in quasiholes, breaking the degeneracy \cite{Nayak08}. In that case, the quasidegenerate states with the smallest total $L$ would be the least affected because they correspond to the states with the quasiholes the furthest away from one another~\cite{jain2007composite,Sitko96}. To test for this effect we compute the energy gap ($\Delta^1_\mathrm{therm}$) between the lowest energy in the continuum and the energy of the quasihole state with the smallest $L$ (not necessarily the smallest energy), i.e., $L = 0$, 1, 0, and 1 for $N = 8$, 10, 12, and 14, respectively, in the thermodynamic limit.  Fig.~\ref{fig-qh-therm} shows this gap is well-defined (note the smaller error bars in the linear extrapolation) and finite in the thermodynamic limit and the smallest $L$ state of the quasidegenerate quasihole states of $H_3$ and $H_2$ remain adiabatically connected.  Our results indicate  the size and physical overlap of quasiholes (for large $L$) likely is leading to a finite-size effect that undermines adiabaticity in our finite-size study of the quasihole gap. 

To further elucidate the above, we study the gap between the lowest energy state in the continuum and the lowest \textit{energy} state of the quasidegenerate manifold of quasihole states ($\Delta^2_\mathrm{therm}$ [not shown]), i.e., $L = 2$, 3, 4, and 5 for $N = 8$, 10, 12, and 14, respectively.  In contrast to $\Delta^1_\mathrm{therm}$,  $\Delta^2_\mathrm{therm}$ is not  well-behaved and eventually decreases and bleeds into  $\delta_\mathrm{therm}$. The error of this extrapolation is also very large indicating  the extrapolation is not particularly linear. This behavior  is consistent with the explanation that the degeneracy is  broken due to interactions between the quasiholes in a finite-sized system caused by the two-body interaction of $H_2$.

The four-quasihole sector at $2Q = 2N-1$ does not show reasonable adiabatic continuity as the degenerate manifold is broken badly by $H_2$ and bleeds into the continuum at finite $\alpha \lesssim 0.5$~\cite{hutzel-thesis}.  From the above discussion of quasiparticle size we conclude that the apparent lack of adiabatic continuity is a reflection of quasihole interactions between closely spaced quasiholes in our finite-sized system studies.  We do find, however, that well-separated quasiholes retain adiabatic continuity in the thermodynamic limit. 
	
\section{particle-hole-symmetry}
\label{sec_phs}

Physical two-body interactions are PH symmetric and so we  expect ground states of two-body Hamiltonians at half-filling to also be PH symmetric in the absence of spontaneous PH symmetry breaking.  In this section we show that, for finite systems sizes on the torus, $\Psi_2$ remains robustly PH symmetric even under PH-symmetry breaking perturbations.  Moreover, we argue that prior evidence for spontaneously broken PH symmetry on the sphere \cite{Peterson08c} can  be interpreted as a expression of an even-odd effect.  We conclude that our numerical calculations do not show unambiguous evidence that $H_2$ spontaneously breaks PH symmetry.

We begin by noting that $\Psi_2$ is nearly identical to the Moore-Read Pfaffian.  (Note  the PH conjugate of $\Psi_2$ compares equally well to the anti-Pfaffian since $H_2$ is PH symmetric.) The numerical wave function overlaps for various system sizes given in Table~\ref{tab:overlaps} quantify how ``identical" $\Psi_2$ is compared to $\Psi_\mathrm{Pf}$.   The overlap is above 0.96 for systems up to $N=16$.  The proximity of these overlaps to unity compares to overlaps found between CF wave functions for odd-denominator FQH states of the form $\nu=n_0/(2pn_0\pm 1)$ and the pure Coulomb ground state in the LLL~\cite{jain2007composite}.

\begin{table}[t!]
\caption{Numerical wave function overlaps between $\Psi_2$ and $\Psi_\mathrm{Pf}$ on the sphere.  Note  the overlaps between $\Psi_2$ at the Moore-Read anti-Pfaffian shift $2Q=2N+1$ and the Moore-Read anti-Pfaffian state $\Psi_\mathrm{aPf}$ are identical to those listed below for $N\rightarrow N-2$.}
\begin{center}
\begin{tabular}{c||c|c|c|c|c}
\hline
 $N$ & 8 & 10 & 12 & 14 & 16 \\
\hline
$\langle \Psi_2 |\Psi_\mathrm{Pf}\rangle$   & 0.9997  & 0.9951 & 0.9869 & 0.9724 & 0.96345 \\
\hline\hline
\end{tabular}
\end{center}
\label{tab:overlaps}
\end{table}%

In the past, it was found that PH symmetrization (on the torus) was able to increase the overlap of $\Psi_\mathrm{Pf}$ with the ground state of the Coulomb interaction for $\nu=5/2$ \cite{Rezayi00}.  Thus, perhaps $\Psi_2$ is identical (or very close) to  the PH symmetrized $\Psi_\mathrm{Pf}$. After all, $\Psi_2$ is the ground state of $H_2=H_3+\overline{H}_3$, with each term producing $\Psi_\mathrm{Pf}$ and $\Psi_\mathrm{aPf}$ independently.   For $N=12$ on the torus the overlap between the PH symmetrized  $\Psi_\mathrm{Pf}$ and $\Psi_2$ is  $0.849$ while the overlaps between $\Psi_2$ and $\Psi_\mathrm{Pf}$ and $\Psi_\mathrm{aPf}$ are both 0.694 (see Appendix~\ref{sec_combining}). So the PH symmetrized $\Psi_\mathrm{Pf}$ is not identical to $\Psi_2$ and we conclude that simple operations acting on $\Psi_\mathrm{Pf}$ (and/or $\Psi_\mathrm{aPf}$) do not generate $\Psi_2$ with high accuracy.

We now consider the possibility of spontaneous PH symmetry breaking of $\Psi_2$ \cite{Peterson08c}.  If $\Psi_\mathrm{Pf}$ explicitly breaks PH symmetry, and the overlap between $\Psi_2$ and $\Psi_\mathrm{Pf}$ is so close to unity, then perhaps $\Psi_2$  spontaneously breaks PH symmetry.  If fact, it has been argued~\cite{Lee07,Levin07} that a ground-state adiabatically connected to $\Psi_\mathrm{Pf}$ must break PH symmetry spontaneously.  Reference~[\onlinecite{Peterson08c}] worked exclusively in the spherical geometry and checking whether $\langle \Psi | \overline{\Psi}\rangle$ is unity, to determine PH symmetry, is not straightforward.  Instead,  the ground state energy of $H_2$ was examined as a function of $N$ (for various flux $2Q$) in the vicinity of half-filling.  The ground state energy of $H_2$ was found to be lower for situations when $2Q=2N-3$ or $2Q=2N+1$ for $N$ even exhibiting a ``wine-bottle" structure.  This was interpreted as evidence for spontaneous PH symmetry breaking.   

However, it is possible  the ``wine-bottle" structure instead points to an even-odd effect.  In Ref.~[\onlinecite{Lu2010}], the ground-state energy per particle of $H_2$ was calculated for $2Q=2N-3$ from $N=4$ to 18 and showed a distinct even-odd effect.  The even-odd effect is consistent with a paired ground state (recall, it has high overlap with the paired Moore-Read Pfaffian state). The even-odd effect in the finite-size calculations could have been the underlying cause of the ``wine-bottle"-shaped energy profile.  Thus, it is unclear if $\Psi_2$ actually spontaneously breaks PH symmetry or is simply a paired state exhibiting an even-odd effect (or both).  
As mentioned above, it is difficult in the spherical geometry to check PH symmetry through the calculation of $\langle \Psi | \overline{\Psi}\rangle$ due to the spherical shift $S$.  The shift, in a sense, explicitly breaks PH symmetry in the basis states themselves (except for the case $S=0$ at $2Q = 2N -1$).  As a result of the shift, taking the PH conjugate of $\Psi_2$ or $\Psi_\mathrm{Pf}$ at  $2Q=2N-3$ changes $N$ at constant flux or changes the flux at constant $N$.  This complicates the analysis~\cite{Pakrouski16}.  In contrast, it is straightforward to compute $\langle \Psi | \overline{\Psi}\rangle$ on the torus  since the shift is zero and the PH conjugate of $\Psi$ has the same quantum numbers as $\Psi$.  If $\Psi$ is PH-symmetric, then $\langle \Psi | \overline{\Psi}\rangle=1$. 
	
The lower two panels of Fig.~\ref{fig-PHSTorus} show $\langle \Psi | \overline{\Psi}\rangle$ for $H(\alpha)$ and $N=8$ and $N=12$.  The data at $H(0)=H_3$ show clearly that $\Psi_\mathrm{Pf}$ breaks PH symmetry, as expected, and at $H(1)=H_2$, $\Psi_2$ is PH symmetric, as expected for the ground state of a two-body Hamiltonian.  Interestingly, for the larger system size ($N=12$) three additional states appear to be separating from the continuum.  Reference~[\onlinecite{Levin07}] pointed out that the degeneracy in the thermodynamic limit of a PH symmetric state in the Pfaffian universality class would theoretically carry a six-fold degeneracy with the extra factor of two arising from PH symmetry (we have already subsumed a factor of two due to center-of-mass).  Our calculation is seemingly the first indication of this effect that is typically absent due to strong finite-size effects.  Importantly,  $\Psi_2$ remains PH-symmetric even with  the explicit PH symmetry breaking term  $(1-\alpha)H_3$ added.  We find that as $\alpha$ is reduced from unity, the additional term does not immediately break PH symmetry and the ground state $\Psi$ of $H(\alpha)$ remains robustly PH symmetric for finite values of $\alpha$.  If the ground state were to spontaneously break PH-symmetry, we would expect $\langle \Psi | \overline{\Psi}\rangle$ to behave qualitatively different as a function of $\alpha$  \cite{Wang2009}.

Before ending this section we briefly consider the ground state energy of $H_2$ at the PH symmetric shift ($S=-1$) in the spherical geometry ($2Q=2N-1$) and ask whether the ground state might be related to the so-called PH-Pfaffian state~\cite{Son15,Bonderson13,Chen14,Milovanovic16,Mishmash18}.  The PH-Pfaffian has  been discussed in relation to recent puzzling experimental results regarding the thermal Hall conductivity at $\nu=5/2$ \cite{Banerjee2018}.  Reference~[\onlinecite{Balram2018}] inadvertently computed the ground state of $H_2$ in the spherical geometry.  Figure~3 and Appendix A of that work studied all even $N$ from $N=8$ to 16.  A uniform ground state with $L=0$ was \emph{only} found for $N=12$ for $H_2$--all other systems were compressible.  These results indicate  the ground state of $H_2$ is not consistent with the PH-Pfaffian. 

To summarize this section, the bottom panels in Fig.~\ref{fig-PHSTorus} suggest that $H_2$ does \emph{not} spontaneously break PH symmetry. This surprising result leads us back to the ``wine-bottle" result of Ref.~[\onlinecite{Peterson08c}].  One interpretation of Ref.~[\onlinecite{Peterson08c}] is that the ``wine-bottle"  merely reflects electron pairing without PH symmetry breaking.  Another possibility is that results presented in Fig.~\ref{fig-PHSTorus} are far from the thermodynamic limit and cannot capture spontaneous PH symmetry breaking.  We therefore conclude  the finite-size results have so far not settled the issue of whether or not $H_2$ spontaneously breaks PH symmetry.

\section{Entanglement properties}
\label{sec_entanglement}

To further study the topological order of $\Psi_2$ we  examine its quantum entanglement properties.  Here we  definitively show that $\Psi_2$ is in the same universality class as the Moore-Read Pfaffian/anti-Pfaffian by investigating the topological entanglement entropy and spectrum.

\subsection{Topological entanglement entropy}	
	
The  topological entanglement entropy of a wave function $\Psi$ is found by  dividing the system into two pieces ($A$ and $B$) and calculating the von Neumann entropy $S_A$ of the partial density matrix of region $A$  by tracing out all   degrees of freedom in $B$ of the total density matrix.  This entropy measures the degree to which the wave function's degrees of freedom in the two subsystems are entangled~\cite{srednicki93}.  A state's  topological order  manifests as a reduction in $S_A$~\cite{kitaev06b,Levin06} as
\BEq
S_A = a L_S - \gamma_\Psi + \mathcal{O}(1/L_S)
\EEq
where $a$ is a nonuniversal constant, $L_S$ is the length of the boundary between the two subsystems, and $\gamma_\Psi>0$ is the topological entanglement entropy. 

We will utilize the spherical geometry for this calculation and consider an orbital partition: geometrically we are partitioning the sphere along a lines of latitude.  The aim is to determine $\gamma_{\Psi_2}$ for $\Psi_2$ compared with the topological entanglement entropy of the Moore-Read Pfaffian,  known exactly to be  $\gamma_{\Psi_\mathrm{Pf}}=\ln\sqrt{8}$~\cite{Moore91,Zozulya07}.   For a topologically ordered state, $\gamma_\Psi=\ln \mathcal{D}$ where $\mathcal{D}=\sqrt{\sum d_i^2}$ is the total quantum dimension and $d_i$ are the quantum dimensions of the quasiparticle excitations of the theory \cite{Nayak08}.  

We follow Zozulya \textit{et al}.~\cite{Zozulya07} in our calculation of $\gamma_{\Psi_2}$ and calculate $S_A$ for various lengths $L_S$ to find the ``$y$-intercept" $\gamma_{\Psi}$.   We focus on systems for which $2Q=2N-3$ and partition along a $z$-component of angular momentum $l_A$.  This orbital partition corresponds to partition at a LLL single-particle state of the form $\eta_{l_A}(z)\sim z^{l_A} \exp{(-|z|^2/4)}$ as the spherical radius is taken to infinity.  The radius of this state is proportional to $\sqrt{l_A}$ so the  boundary length  $L_S\propto\sqrt{l_A}$.  A partition of the sphere along an orbital corresponding to $l_A$ gives an expected  entanglement entropy of the form $S_A = a \sqrt{l_A} - \gamma_{\Psi} + \mathcal{O}(1/\sqrt{l_A})$.  However, the fact that we must calculate for a finite sphere and take the thermodynamic limit makes this procedure more  involved.  In fact, the entanglement entropy of $S_B=S_A$ so, for a given $N$ (and therefore $Q$), we calculate $S_A$ for $l_A=1,\ldots,(2Q+1)/2$ ($l_A = (2Q+1)/2$ is the equator).  Again, following Ref.~[\onlinecite{Zozulya07}] we label $S_A$ for various partitions $l_A$ as $S_{l_A}$ and Table~\ref{ee-table} gives $S_{l_A}(N)$ as a function of $l_A=1,\ldots,(2Q+1)/2$ and $N=6$, 8, 10, 12, and 14 for $\Psi_2$.  One can compare these values to those calculated for the Moore-Read Pfaffian given in Ref.~[\onlinecite{Zozulya07}].  Indeed, the entanglement entropy of $\Psi_2$ is nearly identical to $\Psi_\mathrm{Pf}$.

\begin{table}[t!]
\caption{Entanglement entropy $S_{l_A}$ as a function of $l_A$ for $\Psi_2$ for $N=6$, 8, 10, 12, and 14.  These numbers should be compared to those of the Moore-Read Pfaffian given in Ref.~\onlinecite{Zozulya07}.}
\begin{center}
\begin{tabular}{c|c|c|c|c|c}
\hline
 $l_\mathrm{A}$ & $N=$ 6 & 8 & 10 & 12 & 14 \\
\hline\hline
1 & 0.67301 & 0.68291 & 0.68696 & 0.68901 & 0.69019 \\
2 & 1.15777 & 1.23414 & 1.24039 & 1.24491 & 1.25173 \\
3 & 1.49971 & 1.65389 & 1.68975 & 1.71084 & 1.72946 \\
4 & 1.76712 & 2.03339 & 2.09498 & 2.13745 & 2.17107 \\
5 & \bf{1.88152} & 2.31556 & 2.43072 & 2.50253 & 2.55808 \\
6 &               & 2.48129 & 2.68508 & 2.80127 & 2.88807 \\
7 &               & \bf{2.54204} & 2.86597 & 3.04219 & 3.16802 \\
8 &               &               & 2.97194 & 3.22674 & 3.39942 \\
9 &               &               & \bf{3.00718} & 3.35726 & 3.58560 \\
10 &               &               &               & 3.4354 & 3.72965 \\
11 &               &               &               & \bf{3.4609} & 3.83228 \\
12 &               &               &               &             & 3.89415 \\
13 &               &               &               &             & \bf{3.91462} \\
\hline\hline
\end{tabular}
\end{center}
\label{ee-table}
\end{table}%

\begin{figure}[ht!]
\begin{center}
\includegraphics[width=8cm,angle=0]{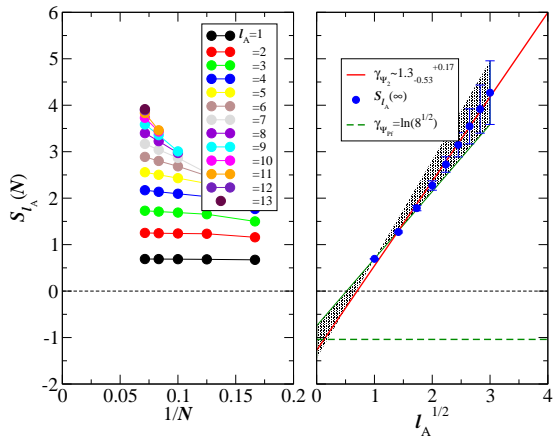}
\caption{(Color online)  Entanglement entropy $S_{l_A}(N)$ versus $1/N$ for various $l_A$ for $\Psi_2$ at $2Q=2N-3$ (left panel).  The right panel shows the thermodynamic limit $S_{l_A}(\infty)$ versus $\sqrt{l_A}$ to determine the topological entanglement entropy $\gamma_{\Psi_2}$.  The error bars are uncertainty of the extrapolation (explained in the text) and the shaded region depicts the upper and lower bounds of the extrapolation.
}
\label{ee-fig}
\end{center}
\end{figure}

To extrapolate to the thermodynamic limit we construct a linear fit of $ S_{l_A}(N)$ versus $1/N$ and the thermodynamic limit ($1/N\rightarrow 0$) is determined (see the left panel of Fig.~\ref{ee-fig}).  Upper and lower bounds of this extrapolation are used to assign an uncertainty of $\pm |S_{l_A}(0) - S_1|$ where $S_1 = S_{l_A}(x_1)(1-x_1/(x_1-x_0) + S_{l_A}(x_0)(x_1/(x_1-x_0)$ where the two smallest values of $N$ are $x_0$ and $x_1$, respectively~\cite{Zozulya07}

A least squares fit of $S_{l_A}(0)$ as a function of $\sqrt{l_A}$ gives $\gamma_{\Psi_2}=1.27$.  To determine the error bars we construct a straight line through the value of $S_{l_A}(0)$ for the largest (smallest) value of $l_A$ plus (minus) $|S_{l_A}(0) - S_1|$ and vice-versa.  We then determine the negative of the $y$-intercept and an upper and lower bound on $\gamma_{\Psi_2}$--these lines are the bounds of the shaded region in the right panel of Fig.~\ref{ee-fig} finally yielding $\gamma_{\Psi_2} = 1.27^{+0.17}_{-0.53}$.
This result contains $\gamma_\mathrm{Pf}=\ln\sqrt{8}\approx 1.04$  within its error bars.  Of course, more work can reduce the error bars in $\gamma_{\Psi_2}$ as was done for $\gamma_{\Psi_\mathrm{Pf}}$ in Ref.~[\onlinecite{Zozulya07}].  Nonetheless, our results show that $\gamma_{\Psi_2}$ is consistent with  $\gamma_{\Psi_\mathrm{Pf}}$.

\subsection{Topological entanglement spectrum}

The entanglement entropy provides limited information about the topological order of a wave function. Consequently, Li and Haldane proposed investigating the entire spectrum of eigenvalues $\{\xi_i\}$ of the reduced density matrix~\cite{Li08}.

\begin{figure}[ht!]
\begin{center}
\includegraphics[width=8cm,angle=0]{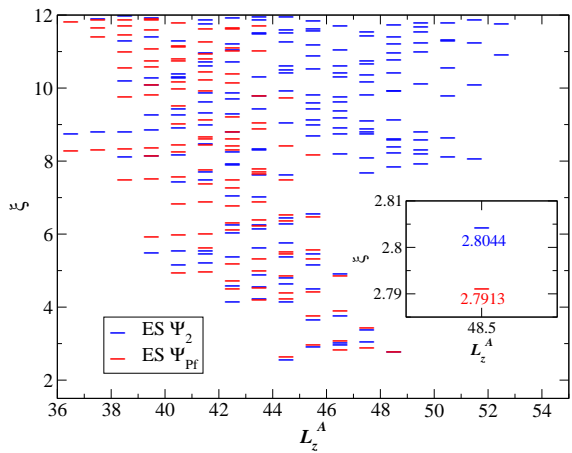}
\caption{(Color online)  Orbital entanglement spectrum for $\Psi_2$ and Moore-Read Pfaffian for $N=14$.  The inset shows the ES of the ``root" configuration\cite{Li08} of both states.  The inset shows a zoom in where the states are barely distinguishable.  The partition is along the spherical equator, i.e., $P[1|1]$ in the notation of Ref.~[\onlinecite{Li08}]. 
}
\label{fig-ES}
\end{center}
\end{figure}

Figure~\ref{fig-ES} shows the orbital entanglement spectrum of $\Psi_2$ and $\Psi_\mathrm{Pf}$ (with which to compare) for $N=14$ particles--we note that smaller system sizes show similar results.  We report the spectra while partitioning the sphere at the equator (in the notation of Li and Haldane~\cite{Li08} this corresponds to $P[0|0]$ for $N/2$ even and $P[1|1]$ for $N/2$ odd).

  The low-lying states of the orbital entanglement spectrum can be used as a ``fingerprint" with which to identify topologically ordered states as they are related to the underlying conformal field theory (CFT) of the corresponding FQH state~\cite{Li08,Chandran11,Qi12}.  The low-lying states of $\Psi_2$ very closely match those of $\Psi_\mathrm{Pf}$ both quantitatively and qualitatively (counting of levels).  There is a clear ``topological entanglement gap"~\cite{Li08} separating the CFT-like low lying states from the higher ``energy" generic non-CFT-like levels.

While the low-lying levels of the entanglement spectrum of $\Psi_2$ match very closely to those of the Moore-Read Pfaffian (or anti-Pfaffian at the anti-Pfaffian shift), the spectrum of $\Psi_2$ has many generic non-CFT-like levels similar, qualitatively, to the second LL Coulomb interaction \cite{Li08,Zhao11,Biddle11}.  From both the entanglement entropy and spectrum it appears that $\Psi_2$ is in the same universality class as the Moore-Read Pfaffian/anti-Pfaffian.

\section{Summary}
\label{sec_summary}

We have studied a physically realistic model for the FQHE in a half-filled LL  to investigate to what extent  PH symmetry breaking and/or the existence of a generating Hamiltonian with three-body terms, crucial to the realization of exotic states that support non-Abelian anyonic excitations in the universality class of the Moore-Read Pfaffian/anti-Pfaffian?

The model we studied, $H_2$, is short ranged and two-body but nonetheless hosts non-Abelian quasihole excitations.  We find  the two-body model has roton and neutral fermion modes as well as quasiparticle pairing as indicated by an even-odd effect and strong overlap with the Moore-Read Pfaffian/anti-Pfaffian.  Furthermore, we find, via entanglement measures,  the  ground state $\Psi_2$ is in the universality class of the Moore-Read Pfaffian/anti-Pfaffian.  

Our most important finding is that the low energy excitations of the Moore-Read Pfaffian are adiabatically connected to those of the  physically realistic two-body model $H_2$.  These excitations include non-Abelian anyons.  Ideally, anyons reside in a topologically protected exactly degenerate manifold. We find, however, a small splitting between these degenerate states likely persists even in the thermodynamic limit, hence, it is possible  the splitting is caused by ``tunneling" between sectors defined by the PH symmetrization operator. While this splitting remains below all energy gaps, further exploration of the cause of this splitting will be useful in establishing the robustness of topological protection in topological quantum computing proposals.

$H_2$ is a physically realistic model which therefore lends itself to quantum state engineering.  For example, efforts to realize the LLL with ultracold atoms \cite{Lewenstein2007,Bloch2008,Dalibard2011,Andre2017} are more likely to be able to engineer $H_2$ than $H_3$.  Further work to realize $H_2$ with ultracold fermions (or Bosonic counterpart models with ultracold bosons) could lead to the exciting possibility of non-Abelian anyons in a tunable environment.

\textit{Acknowledgments}--W. H. \& M.R.P. were supported by the National Science Foundation under Grant No. DMR-1508290, W. H., J. J. M., and M. R. P. acknowledge the the Office of Research and Sponsored Programs at California State University Long Beach, and the W. M. Keck Foundation. M. R. P. acknowledges support in part by the National Science Foundation under Grant No. NSF PHY11-25915.  V.W.S. and P.R. acknowledge support from AFOSR (FA9550-18-1-0505) and ARO (W911NF-16-1-0182).   H. W. is supported by the National Natural Science Foundation of China (NSFC) Grant No. 11474144.  We thank Kiryl Pakrouski for helpful comments on the manuscript.

\appendix
\renewcommand\thefigure{\thesection\arabic{figure}}

\section{Haldane pseudopotential expansion of $H_2$ on the sphere for finite system Sizes}
\label{sec_haldane}

\setcounter{figure}{0}

\begin{figure}[ht!]
\begin{center}
\includegraphics[width=8cm,angle=0]{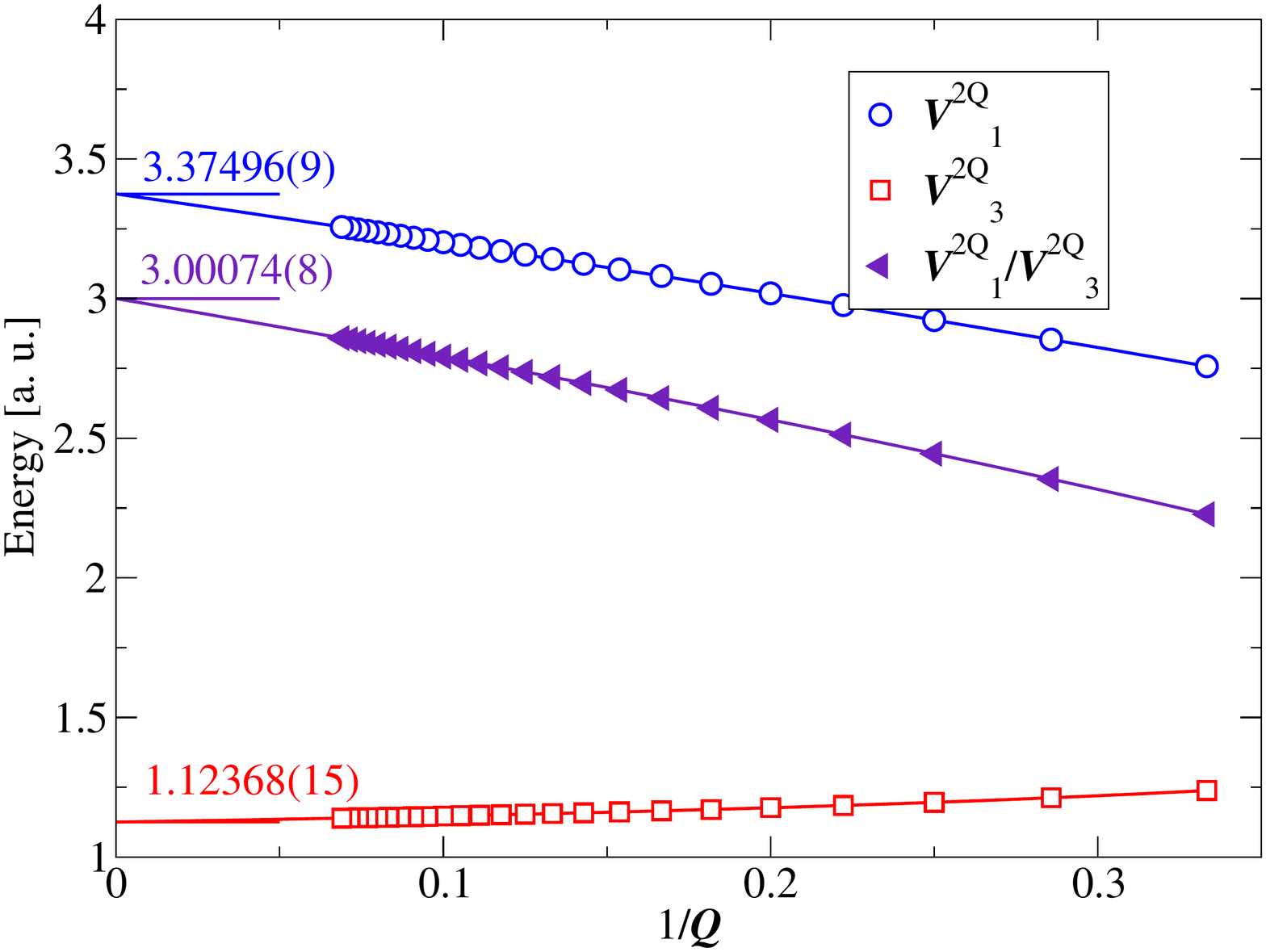}
\caption{(Color online)  The finite-sized Haldane pseudopotentials $V^{2Q}_1$,  $V^{2Q}_3$, and the ratio $V^{2Q}_1/V^{2Q}_3$ of $H_2$ are shown versus the inverse system size ($1/Q$).  The values of $V_1$, $V_3$, and $V_1/V_3$ in the thermodynamic limit  ($1/Q\rightarrow 0$) are shown on the $y$-axis and found via a fourth-order polynomial extrapolation.  The value $V_1/V_3 \rightarrow 3$ is used in the first line of Eq.~(\ref{eq_H_map}).}
\label{fig:vms}
\end{center}
\end{figure}

Any two-body interaction Hamiltonian can be characterized  in terms of Haldane pseudopotentials $V_m$.  The  only nonzero Haldane pseudopotenials for $H_2$ are  $V_1$ and $V_3$ \cite{Peterson08c}. To see this we write $H_3+\overline{H}_3$ and, after cancellation of the three-body terms, construct the remaining finite-size two-body terms on the sphere:
\BEq
H_2 &=&  V^{2Q}_1 \sum_{i<j} \hat{P}_{ij}(1) + V^{2Q}_3\sum_{i<j}\hat{P}_{ij}(3)
\EEq
with all other $V^{2Q}_m=0$ for all even $m$ and for all odd $m>3$ (here $m$ is the pair relative angular momentum in the planar geometry after the appropriate stereographic mapping \cite{Fano86}).  Table~\ref{tab:vms} and Fig.~\ref{fig:vms}  give the  values of $V_1$ and $V_3$, and their ratio $V_1/V_3$, for some relevant system sizes.  Additionally, one can extrapolate the finite-sized spherical pseudopotentials to the thermodynamic limit yielding the pseudopotentials in the planar geometry.  The ratio  $V_1/V_3$ equals three to high precision in the thermodynamic limit.  We note that in Ref.~[\onlinecite{Peterson08c}] the pseudopotential values in the thermodynamic limit were given incorrectly, however, all ratios were correct and all results therein remain unchanged.

Finally, the planar values $V_1 = 3.375$ and $V_3=1.125$ were found  in a very different calculation recently~\cite{Sreejith17} as the ``leading-order" terms in a type of mean-field two-body approximation of a generic three-body interaction term.  

\begin{table}[ht!]
\caption{In the thermodynamic limit $V_1 = 3.37496(9) $, $V_3=1.12368(15)$ and $V_1/V_3=3.00074(8)$.  To a high degree of accuracy, one can calculate $V^{2Q}_1$ and $V^{2Q}_3$ at values of $Q$ in between those given in the table through interpolation--however, you could contact the corresponding author who will be happy to give you values up to around $2Q=32$.}
\begin{center}
\begin{tabular}{c|c|c|c}
\hline
 $2Q$ & $V^{2Q}_{1}$ & $V^{2Q}_3$ & $V^{2Q}_{1}/V^{2Q}_3$ \\
 \hline\hline
9 &  2.976470 & 1.184615 & 2.512605\\ 
13 & 3.104367 & 1.161498 & 2.672727\\
17 & 3.170137 & 1.151297 & 2.753535\\
21 & 3.210124 & 1.145527 & 2.802312\\
25 & 3.237082 & 1.141844 & 2.834960 \\
29 & 3.256420 & 1.139266 & 2.858349 \\
31 & 3.264240 & 1.138269 & 2.867722 \\
\hline\hline
\end{tabular}
\end{center}
\label{tab:vms}
\end{table}%

\section{Tracking States}
\label{sec_tracking}

\begin{figure}[ht!]
\begin{center}
\includegraphics[width=8cm,angle=0]{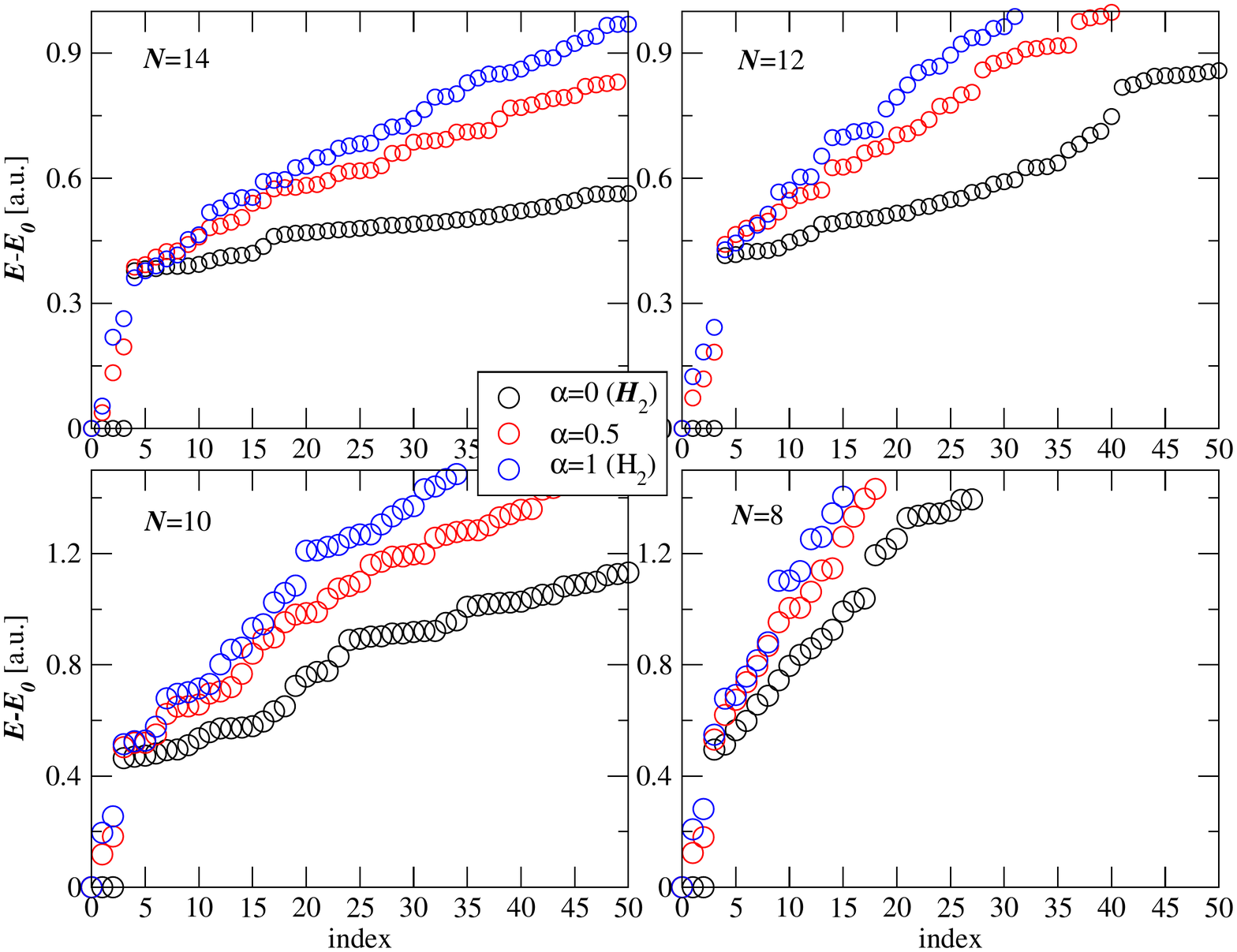}
\caption{(Color online)  The lowest 50 states as of $H(\alpha)$ a function of  ``index" for $N = 8$, 10, 12, and 14 for $\alpha = 0$, 0.5, and 1.   
}
\label{fig-50lowest}
\end{center}
\end{figure}

\begin{figure}[ht!]
\begin{center}
\includegraphics[width=8cm,angle=0]{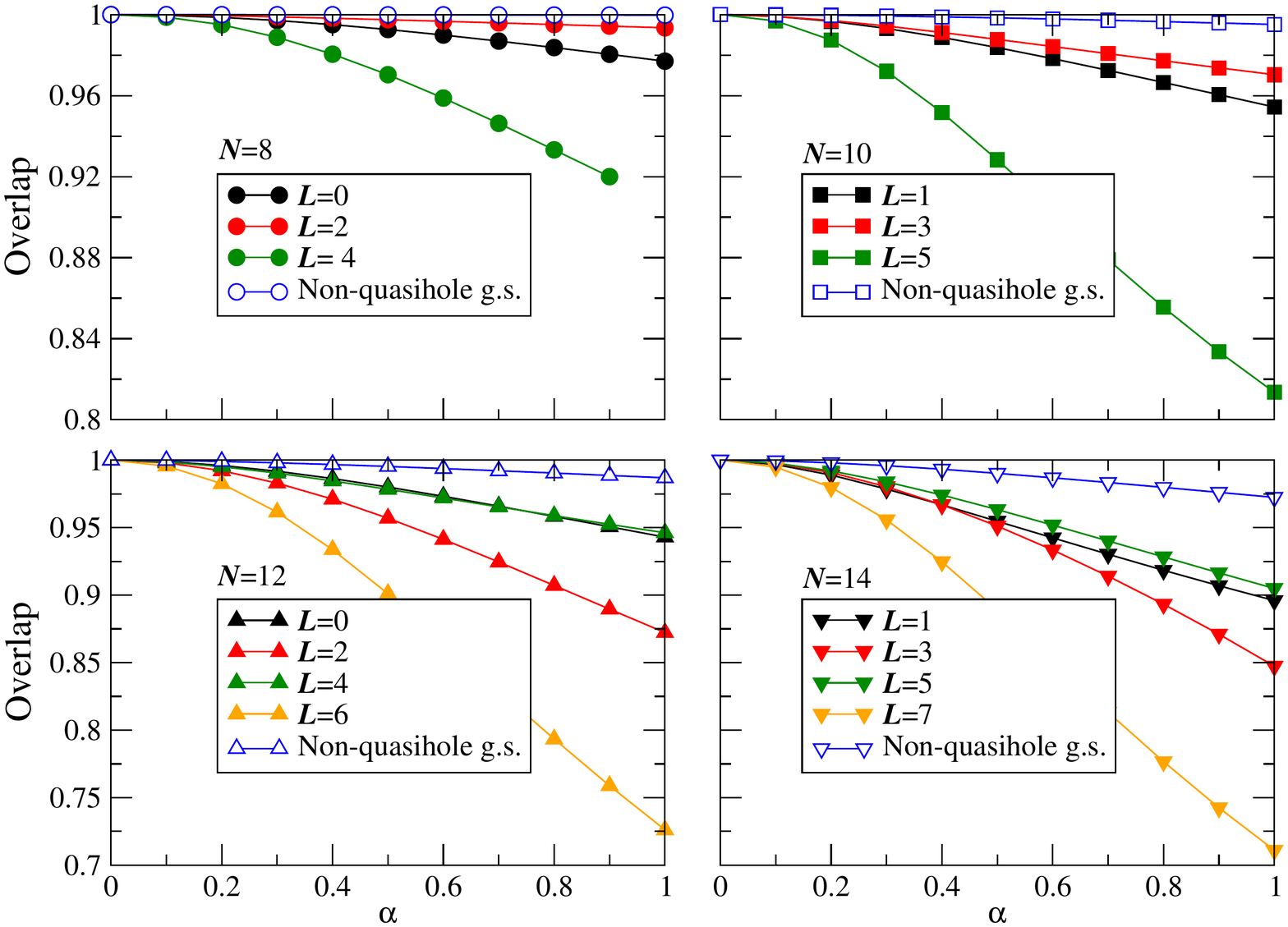}
\caption{(Color online)  The wave function overlap of the exactly degenerate quasihole states, labeled by their angular momentum $L$, of $H_3$ with the quasidegenerate quasihole states of $H(\alpha)$  as a function of $\alpha$ for $N = 8$, 10, 12, and 14. Additionally, the open symbols are the overlaps between the Moore-Read Pfaffian and the ground state of $H(\alpha)$ for the ground state sector. All overlaps are trivially one for $\alpha=0$ and remain large (typically above 0.9 for states with small angular momentum $L$).
}
\label{fig-qh-overlaps}
\end{center}
\end{figure}

In Fig.~\ref{fig-50lowest} we  track individual low-energy states of $H(\alpha)$ more precisely.  The figure shows the lowest 50 states as a function of their ``index"  for $N = 8$, 10, 12, and 14 for $\alpha = 0$, 0.5, and 1. The quasidegenerate states remain inside the gap and adiabatic quasidegeneracy is maintained. More evidence for adiabaticity can be found by calculating the wave function overlap between the quasidegenerate states of $H(\alpha)$ and the exactly degenerate states of $H_3$. In Fig.~\ref{fig-qh-overlaps} we see  the overlaps remain extremely high for each state in the quasidegeneracy. 

\section{Combining Moore-Read Pfaffian and anti-Pfaffian states}
\label{sec_combining}

We can combine the Moore-Read Pfaffian and anti-Pfaffian states to attempt a variational ground state that captures the properties of $\Psi_2$.  We work on the torus to combine these states in an explicit method that effectively PH symmetrizes $\Psi_\mathrm{Pf}$. We first note that $\langle \Psi_\mathrm{Pf}|\Psi_\mathrm{aPf}\rangle\neq 0$ on the torus.  Therefore we use a variant of the Gram-Schmidt procedure (L{\"o}wdin symmetric orthogonalization) to combine the states.  The transformation is
\BEq
|\Psi'_\mathrm{Pf}\rangle &=& c_{+} |\Psi_\mathrm{Pf}\rangle + c_{-} |\Psi_\mathrm{aPf}\rangle\\
|\Psi'_\mathrm{aPf}\rangle &=& c_{-} |\Psi_\mathrm{Pf}\rangle + c_{+} |\Psi_\mathrm{aPf}\rangle
\EEq
where 
\BEq
\hspace{-0.5cm}c_{\pm} &=& \frac{1}{2}\left(\frac{1}{\sqrt{1\mp\langle\Psi_\mathrm{Pf}|\Psi_\mathrm{aPf}\rangle}}\pm\frac{1}{\sqrt{1\pm\langle\Psi_\mathrm{Pf}|\Psi_\mathrm{aPf}\rangle}}\right)
\EEq
Now we see that each state is orthonormal, $\langle \Psi'_\mathrm{Pf}|\Psi'_\mathrm{aPf}\rangle=0$, and the two states are PH conjugates of each other.  Using the above orthogonal states we can construct a PH symmetric state:
\BEq
|\Psi\rangle = \frac{|\Psi'_\mathrm{Pf}\rangle + |\Psi'_\mathrm{aPf}\rangle}{\sqrt{2}}\;.
\EEq
For $N=12$ (aspect ratio $\tau=0.95$) the original overlaps are $\langle \Psi_2 | \Psi_\mathrm{Pf}\rangle = 0.693542 = \langle \Psi_2 | \Psi_\mathrm{aPf}\rangle$ and $\langle\Psi_\mathrm{Pf}|\Psi_\mathrm{aPf}\rangle=0.336155$.  Finally we find that $\langle \Psi_2 | \Psi\rangle = 0.848514$.  Evidently, $\Psi_2$ not identical to a linear combination of $\Psi_\mathrm{Pf}$ and its PH conjugate.


\begin{thebibliography}{85}
\expandafter\ifx\csname natexlab\endcsname\relax\def\natexlab#1{#1}\fi
\expandafter\ifx\csname bibnamefont\endcsname\relax
  \def\bibnamefont#1{#1}\fi
\expandafter\ifx\csname bibfnamefont\endcsname\relax
  \def\bibfnamefont#1{#1}\fi
\expandafter\ifx\csname citenamefont\endcsname\relax
  \def\citenamefont#1{#1}\fi
\expandafter\ifx\csname url\endcsname\relax
  \def\url#1{\texttt{#1}}\fi
\expandafter\ifx\csname urlprefix\endcsname\relax\def\urlprefix{URL }\fi
\providecommand{\bibinfo}[2]{#2}
\providecommand{\eprint}[2][]{\url{#2}}

\bibitem[{\citenamefont{Laughlin}(1983)}]{Laughlin83}
\bibinfo{author}{\bibfnamefont{R.~B.} \bibnamefont{Laughlin}},
  \bibinfo{journal}{Phys. Rev. Lett.} \textbf{\bibinfo{volume}{50}},
  \bibinfo{pages}{1395} (\bibinfo{year}{1983}).

\bibitem[{\citenamefont{Tsui et~al.}(1982)\citenamefont{Tsui, Stormer, and
  Gossard}}]{Tsui82}
\bibinfo{author}{\bibfnamefont{D.~C.} \bibnamefont{Tsui}},
  \bibinfo{author}{\bibfnamefont{H.~L.} \bibnamefont{Stormer}},
  \bibnamefont{and} \bibinfo{author}{\bibfnamefont{A.~C.}
  \bibnamefont{Gossard}}, \bibinfo{journal}{Phys. Rev. Lett.}
  \textbf{\bibinfo{volume}{48}}, \bibinfo{pages}{1559} (\bibinfo{year}{1982}).

\bibitem[{\citenamefont{Haldane}(1983)}]{Haldane83}
\bibinfo{author}{\bibfnamefont{F.~D.~M.} \bibnamefont{Haldane}},
  \bibinfo{journal}{Phys. Rev. Lett.} \textbf{\bibinfo{volume}{51}},
  \bibinfo{pages}{605} (\bibinfo{year}{1983}).

\bibitem[{\citenamefont{Trugman and Kivelson}(1985)}]{Trugman85}
\bibinfo{author}{\bibfnamefont{S.~A.} \bibnamefont{Trugman}} \bibnamefont{and}
  \bibinfo{author}{\bibfnamefont{S.}~\bibnamefont{Kivelson}},
  \bibinfo{journal}{Phys. Rev. B} \textbf{\bibinfo{volume}{31}},
  \bibinfo{pages}{5280} (\bibinfo{year}{1985}).

\bibitem[{\citenamefont{Jain}(1989)}]{Jain89}
\bibinfo{author}{\bibfnamefont{J.~K.} \bibnamefont{Jain}},
  \bibinfo{journal}{Phys. Rev. Lett.} \textbf{\bibinfo{volume}{63}},
  \bibinfo{pages}{199} (\bibinfo{year}{1989}).

\bibitem[{\citenamefont{Jain}(2007)}]{jain2007composite}
\bibinfo{author}{\bibfnamefont{J.}~\bibnamefont{Jain}},
  \emph{\bibinfo{title}{{Composite fermions}}} (\bibinfo{publisher}{Cambridge
  University Press}, \bibinfo{year}{2007}).

\bibitem[{\citenamefont{Halperin et~al.}(1993)\citenamefont{Halperin, Lee, and
  Read}}]{Halperin93}
\bibinfo{author}{\bibfnamefont{B.~I.} \bibnamefont{Halperin}},
  \bibinfo{author}{\bibfnamefont{P.~A.} \bibnamefont{Lee}}, \bibnamefont{and}
  \bibinfo{author}{\bibfnamefont{N.}~\bibnamefont{Read}},
  \bibinfo{journal}{Phys. Rev. B} \textbf{\bibinfo{volume}{47}},
  \bibinfo{pages}{7312} (\bibinfo{year}{1993}).

\bibitem[{\citenamefont{Rezayi and Read}(1994)}]{Rezayi94}
\bibinfo{author}{\bibfnamefont{E.}~\bibnamefont{Rezayi}} \bibnamefont{and}
  \bibinfo{author}{\bibfnamefont{N.}~\bibnamefont{Read}},
  \bibinfo{journal}{Phys. Rev. Lett.} \textbf{\bibinfo{volume}{72}},
  \bibinfo{pages}{900} (\bibinfo{year}{1994}).

\bibitem[{\citenamefont{Rezayi and Haldane}(2000)}]{Rezayi00}
\bibinfo{author}{\bibfnamefont{E.~H.} \bibnamefont{Rezayi}} \bibnamefont{and}
  \bibinfo{author}{\bibfnamefont{F.~D.~M.} \bibnamefont{Haldane}},
  \bibinfo{journal}{Phys. Rev. Lett.} \textbf{\bibinfo{volume}{84}},
  \bibinfo{pages}{4685} (\bibinfo{year}{2000}).

\bibitem[{\citenamefont{Geraedts et~al.}(2016)\citenamefont{Geraedts, Zaletel,
  Mong, Metlitski, Vishwanath, and Motrunich}}]{Geraedts16}
\bibinfo{author}{\bibfnamefont{S.~D.} \bibnamefont{Geraedts}},
  \bibinfo{author}{\bibfnamefont{M.~P.} \bibnamefont{Zaletel}},
  \bibinfo{author}{\bibfnamefont{R.~S.~K.} \bibnamefont{Mong}},
  \bibinfo{author}{\bibfnamefont{M.~A.} \bibnamefont{Metlitski}},
  \bibinfo{author}{\bibfnamefont{A.}~\bibnamefont{Vishwanath}},
  \bibnamefont{and} \bibinfo{author}{\bibfnamefont{O.~I.}
  \bibnamefont{Motrunich}}, \bibinfo{journal}{Science}
  \textbf{\bibinfo{volume}{352}}, \bibinfo{pages}{197} (\bibinfo{year}{2016}).

\bibitem[{\citenamefont{{Willett} et~al.}(1993)\citenamefont{{Willett}, {Ruel},
  {West}, and {Pfeiffer}}}]{Willett93}
\bibinfo{author}{\bibfnamefont{R.~L.} \bibnamefont{{Willett}}},
  \bibinfo{author}{\bibfnamefont{R.~R.} \bibnamefont{{Ruel}}},
  \bibinfo{author}{\bibfnamefont{K.~W.} \bibnamefont{{West}}},
  \bibnamefont{and} \bibinfo{author}{\bibfnamefont{L.~N.}
  \bibnamefont{{Pfeiffer}}}, \bibinfo{journal}{Phys. Rev. Lett.}
  \textbf{\bibinfo{volume}{71}}, \bibinfo{pages}{3846} (\bibinfo{year}{1993}).

\bibitem[{\citenamefont{Willett et~al.}(1987)\citenamefont{Willett, Eisenstein,
  Stormer, Tsui, Gossard, and English}}]{Willett87}
\bibinfo{author}{\bibfnamefont{R.}~\bibnamefont{Willett}},
  \bibinfo{author}{\bibfnamefont{J.~P.} \bibnamefont{Eisenstein}},
  \bibinfo{author}{\bibfnamefont{H.~L.} \bibnamefont{Stormer}},
  \bibinfo{author}{\bibfnamefont{D.~C.} \bibnamefont{Tsui}},
  \bibinfo{author}{\bibfnamefont{A.~C.} \bibnamefont{Gossard}},
  \bibnamefont{and} \bibinfo{author}{\bibfnamefont{J.~H.}
  \bibnamefont{English}}, \bibinfo{journal}{Phys. Rev. Lett.}
  \textbf{\bibinfo{volume}{59}}, \bibinfo{pages}{1776} (\bibinfo{year}{1987}).

\bibitem[{\citenamefont{Moore and Read}(1991)}]{Moore91}
\bibinfo{author}{\bibfnamefont{G.}~\bibnamefont{Moore}} \bibnamefont{and}
  \bibinfo{author}{\bibfnamefont{N.}~\bibnamefont{Read}},
  \bibinfo{journal}{Nucl. Phys. B} \textbf{\bibinfo{volume}{360}},
  \bibinfo{pages}{362} (\bibinfo{year}{1991}).

\bibitem[{\citenamefont{Scarola et~al.}(2000)\citenamefont{Scarola, Park, and
  Jain}}]{Scarola00}
\bibinfo{author}{\bibfnamefont{V.~W.} \bibnamefont{Scarola}},
  \bibinfo{author}{\bibfnamefont{K.}~\bibnamefont{Park}}, \bibnamefont{and}
  \bibinfo{author}{\bibfnamefont{J.~K.} \bibnamefont{Jain}},
  \bibinfo{journal}{Nature} \textbf{\bibinfo{volume}{406}}, \bibinfo{pages}{863
  EP } (\bibinfo{year}{2000}).

\bibitem[{\citenamefont{M\"oller and Simon}(2008)}]{Moller2008}
\bibinfo{author}{\bibfnamefont{G.}~\bibnamefont{M\"oller}} \bibnamefont{and}
  \bibinfo{author}{\bibfnamefont{S.~H.} \bibnamefont{Simon}},
  \bibinfo{journal}{Phys. Rev. B} \textbf{\bibinfo{volume}{77}},
  \bibinfo{pages}{075319} (\bibinfo{year}{2008}).

\bibitem[{\citenamefont{Read and Green}(2000)}]{Read00}
\bibinfo{author}{\bibfnamefont{N.}~\bibnamefont{Read}} \bibnamefont{and}
  \bibinfo{author}{\bibfnamefont{D.}~\bibnamefont{Green}},
  \bibinfo{journal}{Phys. Rev. B} \textbf{\bibinfo{volume}{61}},
  \bibinfo{pages}{10267} (\bibinfo{year}{2000}).

\bibitem[{\citenamefont{Morf}(1998)}]{Morf98}
\bibinfo{author}{\bibfnamefont{R.~H.} \bibnamefont{Morf}},
  \bibinfo{journal}{Phys. Rev. Lett.} \textbf{\bibinfo{volume}{80}},
  \bibinfo{pages}{1505} (\bibinfo{year}{1998}).

\bibitem[{\citenamefont{Peterson
  et~al.}(2008{\natexlab{a}})\citenamefont{Peterson, Jolicoeur, and
  Das~Sarma}}]{Peterson08}
\bibinfo{author}{\bibfnamefont{M.~R.} \bibnamefont{Peterson}},
  \bibinfo{author}{\bibfnamefont{T.}~\bibnamefont{Jolicoeur}},
  \bibnamefont{and}
  \bibinfo{author}{\bibfnamefont{S.}~\bibnamefont{Das~Sarma}},
  \bibinfo{journal}{Phys. Rev. Lett.} \textbf{\bibinfo{volume}{101}},
  \bibinfo{pages}{016807} (\bibinfo{year}{2008}{\natexlab{a}}).

\bibitem[{\citenamefont{Peterson
  et~al.}(2008{\natexlab{b}})\citenamefont{Peterson, Jolicoeur, and
  Das~Sarma}}]{Peterson08b}
\bibinfo{author}{\bibfnamefont{M.~R.} \bibnamefont{Peterson}},
  \bibinfo{author}{\bibfnamefont{T.}~\bibnamefont{Jolicoeur}},
  \bibnamefont{and}
  \bibinfo{author}{\bibfnamefont{S.}~\bibnamefont{Das~Sarma}},
  \bibinfo{journal}{Phys. Rev. B} \textbf{\bibinfo{volume}{78}},
  \bibinfo{pages}{155308} (\bibinfo{year}{2008}{\natexlab{b}}).

\bibitem[{\citenamefont{Feiguin et~al.}(2008)\citenamefont{Feiguin, Rezayi,
  Nayak, and Das~Sarma}}]{Feiguin08}
\bibinfo{author}{\bibfnamefont{A.~E.} \bibnamefont{Feiguin}},
  \bibinfo{author}{\bibfnamefont{E.}~\bibnamefont{Rezayi}},
  \bibinfo{author}{\bibfnamefont{C.}~\bibnamefont{Nayak}}, \bibnamefont{and}
  \bibinfo{author}{\bibfnamefont{S.}~\bibnamefont{Das~Sarma}},
  \bibinfo{journal}{Phys. Rev. Lett.} \textbf{\bibinfo{volume}{100}},
  \bibinfo{pages}{166803} (\bibinfo{year}{2008}).

\bibitem[{\citenamefont{Feiguin et~al.}(2009)\citenamefont{Feiguin, Rezayi,
  Yang, Nayak, and Das~Sarma}}]{Feiguin09}
\bibinfo{author}{\bibfnamefont{A.~E.} \bibnamefont{Feiguin}},
  \bibinfo{author}{\bibfnamefont{E.}~\bibnamefont{Rezayi}},
  \bibinfo{author}{\bibfnamefont{K.}~\bibnamefont{Yang}},
  \bibinfo{author}{\bibfnamefont{C.}~\bibnamefont{Nayak}}, \bibnamefont{and}
  \bibinfo{author}{\bibfnamefont{S.}~\bibnamefont{Das~Sarma}},
  \bibinfo{journal}{Phys. Rev. B} \textbf{\bibinfo{volume}{79}},
  \bibinfo{pages}{115322} (\bibinfo{year}{2009}).

\bibitem[{\citenamefont{W\'ojs et~al.}(2010)\citenamefont{W\'ojs,
  T\ifmmode~\mbox{\H{o}}\else \H{o}\fi{}ke, and Jain}}]{Wojs10}
\bibinfo{author}{\bibfnamefont{A.}~\bibnamefont{W\'ojs}},
  \bibinfo{author}{\bibfnamefont{C.}~\bibnamefont{T\ifmmode~\mbox{\H{o}}\else
  \H{o}\fi{}ke}}, \bibnamefont{and} \bibinfo{author}{\bibfnamefont{J.~K.}
  \bibnamefont{Jain}}, \bibinfo{journal}{Phys. Rev. Lett.}
  \textbf{\bibinfo{volume}{105}}, \bibinfo{pages}{096802}
  (\bibinfo{year}{2010}).

\bibitem[{\citenamefont{{Storni} et~al.}(2010)\citenamefont{{Storni}, {Morf},
  and {Das Sarma}}}]{Storni10}
\bibinfo{author}{\bibfnamefont{M.}~\bibnamefont{{Storni}}},
  \bibinfo{author}{\bibfnamefont{R.~H.} \bibnamefont{{Morf}}},
  \bibnamefont{and} \bibinfo{author}{\bibfnamefont{S.}~\bibnamefont{{Das
  Sarma}}}, \bibinfo{journal}{Phys. Rev. Lett.} \textbf{\bibinfo{volume}{104}},
  \bibinfo{pages}{076803} (\bibinfo{year}{2010}).

\bibitem[{\citenamefont{Rezayi and Simon}(2011)}]{Rezayi2011}
\bibinfo{author}{\bibfnamefont{E.~H.} \bibnamefont{Rezayi}} \bibnamefont{and}
  \bibinfo{author}{\bibfnamefont{S.~H.} \bibnamefont{Simon}},
  \bibinfo{journal}{Phys. Rev. Lett.} \textbf{\bibinfo{volume}{106}},
  \bibinfo{pages}{116801} (\bibinfo{year}{2011}).

\bibitem[{\citenamefont{Pakrouski et~al.}(2015)\citenamefont{Pakrouski,
  Peterson, Jolicoeur, Scarola, Nayak, and Troyer}}]{Pakrouski2015}
\bibinfo{author}{\bibfnamefont{K.}~\bibnamefont{Pakrouski}},
  \bibinfo{author}{\bibfnamefont{M.~R.} \bibnamefont{Peterson}},
  \bibinfo{author}{\bibfnamefont{T.}~\bibnamefont{Jolicoeur}},
  \bibinfo{author}{\bibfnamefont{V.~W.} \bibnamefont{Scarola}},
  \bibinfo{author}{\bibfnamefont{C.}~\bibnamefont{Nayak}}, \bibnamefont{and}
  \bibinfo{author}{\bibfnamefont{M.}~\bibnamefont{Troyer}},
  \bibinfo{journal}{Phys. Rev. X} \textbf{\bibinfo{volume}{5}},
  \bibinfo{pages}{021004} (\bibinfo{year}{2015}).

\bibitem[{\citenamefont{Zaletel et~al.}(2015)\citenamefont{Zaletel, Mong,
  Pollmann, and Rezayi}}]{Zaletel15}
\bibinfo{author}{\bibfnamefont{M.~P.} \bibnamefont{Zaletel}},
  \bibinfo{author}{\bibfnamefont{R.~S.~K.} \bibnamefont{Mong}},
  \bibinfo{author}{\bibfnamefont{F.}~\bibnamefont{Pollmann}}, \bibnamefont{and}
  \bibinfo{author}{\bibfnamefont{E.~H.} \bibnamefont{Rezayi}},
  \bibinfo{journal}{Phys. Rev. B} \textbf{\bibinfo{volume}{91}},
  \bibinfo{pages}{045115} (\bibinfo{year}{2015}).

\bibitem[{\citenamefont{Tylan-Tyler and Lyanda-Geller}(2015)}]{Tylan2015}
\bibinfo{author}{\bibfnamefont{A.}~\bibnamefont{Tylan-Tyler}} \bibnamefont{and}
  \bibinfo{author}{\bibfnamefont{Y.}~\bibnamefont{Lyanda-Geller}},
  \bibinfo{journal}{Phys. Rev. B} \textbf{\bibinfo{volume}{91}},
  \bibinfo{pages}{205404} (\bibinfo{year}{2015}).

\bibitem[{\citenamefont{Rezayi}(2017)}]{Rezayi2017}
\bibinfo{author}{\bibfnamefont{E.~H.} \bibnamefont{Rezayi}},
  \bibinfo{journal}{Phys. Rev. Lett.} \textbf{\bibinfo{volume}{119}},
  \bibinfo{pages}{026801} (\bibinfo{year}{2017}).

\bibitem[{\citenamefont{Nayak and Wilczek}(1996)}]{Nayak96c}
\bibinfo{author}{\bibfnamefont{C.}~\bibnamefont{Nayak}} \bibnamefont{and}
  \bibinfo{author}{\bibfnamefont{F.}~\bibnamefont{Wilczek}},
  \bibinfo{journal}{Nucl. Phys. B} \textbf{\bibinfo{volume}{479}},
  \bibinfo{pages}{529} (\bibinfo{year}{1996}).

\bibitem[{\citenamefont{Tserkovnyak and Simon}(2003)}]{Tserkovnyak03}
\bibinfo{author}{\bibfnamefont{Y.}~\bibnamefont{Tserkovnyak}} \bibnamefont{and}
  \bibinfo{author}{\bibfnamefont{S.~H.} \bibnamefont{Simon}},
  \bibinfo{journal}{Phys. Rev. Lett.} \textbf{\bibinfo{volume}{90}},
  \bibinfo{pages}{016802} (\bibinfo{year}{2003}).

\bibitem[{\citenamefont{Baraban et~al.}(2009)\citenamefont{Baraban, Zikos,
  Bonesteel, and Simon}}]{Baraban09}
\bibinfo{author}{\bibfnamefont{M.}~\bibnamefont{Baraban}},
  \bibinfo{author}{\bibfnamefont{G.}~\bibnamefont{Zikos}},
  \bibinfo{author}{\bibfnamefont{N.}~\bibnamefont{Bonesteel}},
  \bibnamefont{and} \bibinfo{author}{\bibfnamefont{S.~H.} \bibnamefont{Simon}},
  \bibinfo{journal}{Phys. Rev. Lett.} \textbf{\bibinfo{volume}{103}},
  \bibinfo{pages}{076801} (\bibinfo{year}{2009}).

\bibitem[{\citenamefont{Bonderson
  et~al.}(2011{\natexlab{a}})\citenamefont{Bonderson, Gurarie, and
  Nayak}}]{Bonderson11a}
\bibinfo{author}{\bibfnamefont{P.}~\bibnamefont{Bonderson}},
  \bibinfo{author}{\bibfnamefont{V.}~\bibnamefont{Gurarie}}, \bibnamefont{and}
  \bibinfo{author}{\bibfnamefont{C.}~\bibnamefont{Nayak}},
  \bibinfo{journal}{Phys. Rev. B} \textbf{\bibinfo{volume}{83}},
  \bibinfo{pages}{075303} (\bibinfo{year}{2011}{\natexlab{a}}).

\bibitem[{\citenamefont{Das~Sarma et~al.}(2005)\citenamefont{Das~Sarma,
  Freedman, and Nayak}}]{DasSarma05}
\bibinfo{author}{\bibfnamefont{S.}~\bibnamefont{Das~Sarma}},
  \bibinfo{author}{\bibfnamefont{M.}~\bibnamefont{Freedman}}, \bibnamefont{and}
  \bibinfo{author}{\bibfnamefont{C.}~\bibnamefont{Nayak}},
  \bibinfo{journal}{Phys. Rev. Lett.} \textbf{\bibinfo{volume}{94}},
  \bibinfo{pages}{166802} (\bibinfo{year}{2005}).

\bibitem[{\citenamefont{Nayak et~al.}(2008)\citenamefont{Nayak, Simon, Stern,
  Freedman, and Das~Sarma}}]{Nayak08}
\bibinfo{author}{\bibfnamefont{C.}~\bibnamefont{Nayak}},
  \bibinfo{author}{\bibfnamefont{S.~H.} \bibnamefont{Simon}},
  \bibinfo{author}{\bibfnamefont{A.}~\bibnamefont{Stern}},
  \bibinfo{author}{\bibfnamefont{M.}~\bibnamefont{Freedman}}, \bibnamefont{and}
  \bibinfo{author}{\bibfnamefont{S.}~\bibnamefont{Das~Sarma}},
  \bibinfo{journal}{Rev. Mod. Phys.} \textbf{\bibinfo{volume}{80}},
  \bibinfo{pages}{1083} (\bibinfo{year}{2008}).

\bibitem[{\citenamefont{Greiter et~al.}(1991)\citenamefont{Greiter, Wen, and
  Wilczek}}]{Greiter91}
\bibinfo{author}{\bibfnamefont{M.}~\bibnamefont{Greiter}},
  \bibinfo{author}{\bibfnamefont{X.-G.} \bibnamefont{Wen}}, \bibnamefont{and}
  \bibinfo{author}{\bibfnamefont{F.}~\bibnamefont{Wilczek}},
  \bibinfo{journal}{Phys. Rev. Lett.} \textbf{\bibinfo{volume}{66}},
  \bibinfo{pages}{3205} (\bibinfo{year}{1991}).

\bibitem[{\citenamefont{Read and Rezayi}(1996)}]{Read96}
\bibinfo{author}{\bibfnamefont{N.}~\bibnamefont{Read}} \bibnamefont{and}
  \bibinfo{author}{\bibfnamefont{E.}~\bibnamefont{Rezayi}},
  \bibinfo{journal}{Phys. Rev. B} \textbf{\bibinfo{volume}{54}},
  \bibinfo{pages}{16864} (\bibinfo{year}{1996}).

\bibitem[{\citenamefont{Wang et~al.}(2009)\citenamefont{Wang, Sheng, and
  Haldane}}]{Wang2009}
\bibinfo{author}{\bibfnamefont{H.}~\bibnamefont{Wang}},
  \bibinfo{author}{\bibfnamefont{D.~N.} \bibnamefont{Sheng}}, \bibnamefont{and}
  \bibinfo{author}{\bibfnamefont{F.~D.~M.} \bibnamefont{Haldane}},
  \bibinfo{journal}{Phys. Rev. B} \textbf{\bibinfo{volume}{80}},
  \bibinfo{pages}{241311} (\bibinfo{year}{2009}).

\bibitem[{\citenamefont{Bishara and Nayak}(2009)}]{Bishara09a}
\bibinfo{author}{\bibfnamefont{W.}~\bibnamefont{Bishara}} \bibnamefont{and}
  \bibinfo{author}{\bibfnamefont{C.}~\bibnamefont{Nayak}},
  \bibinfo{journal}{Phys. Rev. B} \textbf{\bibinfo{volume}{80}},
  \bibinfo{pages}{121302} (\bibinfo{year}{2009}).

\bibitem[{\citenamefont{Peterson and Nayak}(2013)}]{Peterson13b}
\bibinfo{author}{\bibfnamefont{M.~R.} \bibnamefont{Peterson}} \bibnamefont{and}
  \bibinfo{author}{\bibfnamefont{C.}~\bibnamefont{Nayak}},
  \bibinfo{journal}{Phys. Rev. B} \textbf{\bibinfo{volume}{87}},
  \bibinfo{pages}{245129} (\bibinfo{year}{2013}).

\bibitem[{\citenamefont{Sodemann and MacDonald}(2013)}]{Sodemann2013}
\bibinfo{author}{\bibfnamefont{I.}~\bibnamefont{Sodemann}} \bibnamefont{and}
  \bibinfo{author}{\bibfnamefont{A.~H.} \bibnamefont{MacDonald}},
  \bibinfo{journal}{Phys. Rev. B} \textbf{\bibinfo{volume}{87}},
  \bibinfo{pages}{245425} (\bibinfo{year}{2013}).

\bibitem[{\citenamefont{Simon and Rezayi}(2013)}]{Rezayi2013}
\bibinfo{author}{\bibfnamefont{S.~H.} \bibnamefont{Simon}} \bibnamefont{and}
  \bibinfo{author}{\bibfnamefont{E.~H.} \bibnamefont{Rezayi}},
  \bibinfo{journal}{Phys. Rev. B} \textbf{\bibinfo{volume}{87}},
  \bibinfo{pages}{155426} (\bibinfo{year}{2013}).

\bibitem[{\citenamefont{Wooten et~al.}(2013)\citenamefont{Wooten, Macek, and
  Quinn}}]{Wooten13}
\bibinfo{author}{\bibfnamefont{R.~E.} \bibnamefont{Wooten}},
  \bibinfo{author}{\bibfnamefont{J.~H.} \bibnamefont{Macek}}, \bibnamefont{and}
  \bibinfo{author}{\bibfnamefont{J.~J.} \bibnamefont{Quinn}},
  \bibinfo{journal}{Phys. Rev. B} \textbf{\bibinfo{volume}{88}},
  \bibinfo{pages}{155421} (\bibinfo{year}{2013}).

\bibitem[{\citenamefont{Toke and Jain}(2006)}]{Toke06a}
\bibinfo{author}{\bibfnamefont{C.}~\bibnamefont{Toke}} \bibnamefont{and}
  \bibinfo{author}{\bibfnamefont{J.~K.} \bibnamefont{Jain}},
  \bibinfo{journal}{Phys. Rev. Lett.} \textbf{\bibinfo{volume}{96}},
  \bibinfo{pages}{246805} (\bibinfo{year}{2006}).

\bibitem[{\citenamefont{Peterson
  et~al.}(2008{\natexlab{c}})\citenamefont{Peterson, Park, and {Das
  Sarma}}}]{Peterson08c}
\bibinfo{author}{\bibfnamefont{M.~R.} \bibnamefont{Peterson}},
  \bibinfo{author}{\bibfnamefont{K.}~\bibnamefont{Park}}, \bibnamefont{and}
  \bibinfo{author}{\bibfnamefont{S.}~\bibnamefont{{Das Sarma}}},
  \bibinfo{journal}{Phys. Rev. Lett.} \textbf{\bibinfo{volume}{101}},
  \bibinfo{pages}{156803} (\bibinfo{year}{2008}{\natexlab{c}}).

\bibitem[{\citenamefont{Lee et~al.}(2007)\citenamefont{Lee, Ryu, Nayak, and
  Fisher}}]{Lee07}
\bibinfo{author}{\bibfnamefont{S.-S.} \bibnamefont{Lee}},
  \bibinfo{author}{\bibfnamefont{S.}~\bibnamefont{Ryu}},
  \bibinfo{author}{\bibfnamefont{C.}~\bibnamefont{Nayak}}, \bibnamefont{and}
  \bibinfo{author}{\bibfnamefont{M.~P.~A.} \bibnamefont{Fisher}},
  \bibinfo{journal}{Phys. Rev. Lett.} \textbf{\bibinfo{volume}{99}},
  \bibinfo{pages}{236807} (\bibinfo{year}{2007}).

\bibitem[{\citenamefont{Levin et~al.}(2007)\citenamefont{Levin, Halperin, and
  Rosenow}}]{Levin07}
\bibinfo{author}{\bibfnamefont{M.}~\bibnamefont{Levin}},
  \bibinfo{author}{\bibfnamefont{B.~I.} \bibnamefont{Halperin}},
  \bibnamefont{and} \bibinfo{author}{\bibfnamefont{B.}~\bibnamefont{Rosenow}},
  \bibinfo{journal}{Phys. Rev. Lett.} \textbf{\bibinfo{volume}{99}},
  \bibinfo{pages}{236806} (\bibinfo{year}{2007}).

\bibitem[{\citenamefont{Lu et~al.}(2010)\citenamefont{Lu, Das~Sarma, and
  Park}}]{Lu2010}
\bibinfo{author}{\bibfnamefont{H.}~\bibnamefont{Lu}},
  \bibinfo{author}{\bibfnamefont{S.}~\bibnamefont{Das~Sarma}},
  \bibnamefont{and} \bibinfo{author}{\bibfnamefont{K.}~\bibnamefont{Park}},
  \bibinfo{journal}{Phys. Rev. B} \textbf{\bibinfo{volume}{82}},
  \bibinfo{pages}{201303} (\bibinfo{year}{2010}).

\bibitem[{\citenamefont{M\"oller et~al.}(2011)\citenamefont{M\"oller, W\'ojs,
  and Cooper}}]{Moller2011}
\bibinfo{author}{\bibfnamefont{G.}~\bibnamefont{M\"oller}},
  \bibinfo{author}{\bibfnamefont{A.}~\bibnamefont{W\'ojs}}, \bibnamefont{and}
  \bibinfo{author}{\bibfnamefont{N.~R.} \bibnamefont{Cooper}},
  \bibinfo{journal}{Phys. Rev. Lett.} \textbf{\bibinfo{volume}{107}},
  \bibinfo{pages}{036803} (\bibinfo{year}{2011}).

\bibitem[{\citenamefont{Wurstbauer et~al.}(2013)\citenamefont{Wurstbauer, West,
  Pfeiffer, and Pinczuk}}]{Wurstbauer2013}
\bibinfo{author}{\bibfnamefont{U.}~\bibnamefont{Wurstbauer}},
  \bibinfo{author}{\bibfnamefont{K.~W.} \bibnamefont{West}},
  \bibinfo{author}{\bibfnamefont{L.~N.} \bibnamefont{Pfeiffer}},
  \bibnamefont{and} \bibinfo{author}{\bibfnamefont{A.}~\bibnamefont{Pinczuk}},
  \bibinfo{journal}{Phys. Rev. Lett.} \textbf{\bibinfo{volume}{110}},
  \bibinfo{pages}{026801} (\bibinfo{year}{2013}).

\bibitem[{foo()}]{footnote}
\bibinfo{note}{Numerical studies find the ground state of the realistic Coulomb
  interaction at $\nu=5/2$ is fully spin-polarized
  \cite{Morf98,Dimov08,Feiguin09,Biddle13}. Experiments, however, are somewhat
  mixed \cite{Stern10,Rhone10,Tiemann828}.}

\bibitem[{\citenamefont{Kohn and Luttinger}(1965)}]{Kohn65}
\bibinfo{author}{\bibfnamefont{W.}~\bibnamefont{Kohn}} \bibnamefont{and}
  \bibinfo{author}{\bibfnamefont{J.~M.} \bibnamefont{Luttinger}},
  \bibinfo{journal}{Phys. Rev. Lett.} \textbf{\bibinfo{volume}{15}},
  \bibinfo{pages}{524} (\bibinfo{year}{1965}).

\bibitem[{\citenamefont{Chubukov}(1993)}]{Chubukov93}
\bibinfo{author}{\bibfnamefont{A.~V.} \bibnamefont{Chubukov}},
  \bibinfo{journal}{Phys. Rev. B} \textbf{\bibinfo{volume}{48}},
  \bibinfo{pages}{1097} (\bibinfo{year}{1993}).

\bibitem[{\citenamefont{Wen and Niu}(1990)}]{Wen90b}
\bibinfo{author}{\bibfnamefont{X.~G.} \bibnamefont{Wen}} \bibnamefont{and}
  \bibinfo{author}{\bibfnamefont{Q.}~\bibnamefont{Niu}},
  \bibinfo{journal}{Phys. Rev. B} \textbf{\bibinfo{volume}{41}},
  \bibinfo{pages}{9377} (\bibinfo{year}{1990}).

\bibitem[{\citenamefont{Bonderson
  et~al.}(2011{\natexlab{b}})\citenamefont{Bonderson, Feiguin, and
  Nayak}}]{Bonderson11c}
\bibinfo{author}{\bibfnamefont{P.}~\bibnamefont{Bonderson}},
  \bibinfo{author}{\bibfnamefont{A.~E.} \bibnamefont{Feiguin}},
  \bibnamefont{and} \bibinfo{author}{\bibfnamefont{C.}~\bibnamefont{Nayak}},
  \bibinfo{journal}{Phys. Rev. Lett.} \textbf{\bibinfo{volume}{106}},
  \bibinfo{pages}{186802} (\bibinfo{year}{2011}{\natexlab{b}}).

\bibitem[{\citenamefont{Greiter et~al.}(1992)\citenamefont{Greiter, Wen, and
  Wilczek}}]{Greiter92}
\bibinfo{author}{\bibfnamefont{M.}~\bibnamefont{Greiter}},
  \bibinfo{author}{\bibfnamefont{X.~G.} \bibnamefont{Wen}}, \bibnamefont{and}
  \bibinfo{author}{\bibfnamefont{F.}~\bibnamefont{Wilczek}},
  \bibinfo{journal}{Nucl. Phys. B} \textbf{\bibinfo{volume}{374}},
  \bibinfo{pages}{567} (\bibinfo{year}{1992}).

\bibitem[{\citenamefont{Sitko et~al.}(1996)\citenamefont{Sitko, Yi, Yi, and
  Quinn}}]{Sitko96}
\bibinfo{author}{\bibfnamefont{P.}~\bibnamefont{Sitko}},
  \bibinfo{author}{\bibfnamefont{S.~N.} \bibnamefont{Yi}},
  \bibinfo{author}{\bibfnamefont{K.~S.} \bibnamefont{Yi}}, \bibnamefont{and}
  \bibinfo{author}{\bibfnamefont{J.~J.} \bibnamefont{Quinn}},
  \bibinfo{journal}{Phys. Rev. Lett.} \textbf{\bibinfo{volume}{76}},
  \bibinfo{pages}{3396} (\bibinfo{year}{1996}).

\bibitem[{\citenamefont{Hutzel}(2018)}]{hutzel-thesis}
\bibinfo{author}{\bibfnamefont{W.}~\bibnamefont{Hutzel}}
, \bibinfo{note}{{M}.{S}c. thesis, California State University Long Beach, 2018}.

\bibitem[{\citenamefont{Pakrouski et~al.}(2016)\citenamefont{Pakrouski, Troyer,
  Wu, Das~Sarma, and Peterson}}]{Pakrouski16}
\bibinfo{author}{\bibfnamefont{K.}~\bibnamefont{Pakrouski}},
  \bibinfo{author}{\bibfnamefont{M.}~\bibnamefont{Troyer}},
  \bibinfo{author}{\bibfnamefont{Y.-L.} \bibnamefont{Wu}},
  \bibinfo{author}{\bibfnamefont{S.}~\bibnamefont{Das~Sarma}},
  \bibnamefont{and} \bibinfo{author}{\bibfnamefont{M.~R.}
  \bibnamefont{Peterson}}, \bibinfo{journal}{Phys. Rev. B}
  \textbf{\bibinfo{volume}{94}}, \bibinfo{pages}{075108}
  (\bibinfo{year}{2016}).

\bibitem[{\citenamefont{Son}(2015)}]{Son15}
\bibinfo{author}{\bibfnamefont{D.~T.} \bibnamefont{Son}},
  \bibinfo{journal}{Phys. Rev. X} \textbf{\bibinfo{volume}{5}},
  \bibinfo{pages}{031027} (\bibinfo{year}{2015}).

\bibitem[{\citenamefont{Bonderson et~al.}(2013)\citenamefont{Bonderson, Nayak,
  and Qi}}]{Bonderson13}
\bibinfo{author}{\bibfnamefont{P.}~\bibnamefont{Bonderson}},
  \bibinfo{author}{\bibfnamefont{C.}~\bibnamefont{Nayak}}, \bibnamefont{and}
  \bibinfo{author}{\bibfnamefont{X.-L.} \bibnamefont{Qi}},
  \bibinfo{journal}{Journal of Statistical Mechanics: Theory and Experiment}
  \textbf{\bibinfo{volume}{2013}}, \bibinfo{pages}{P09016}
  (\bibinfo{year}{2013}).

\bibitem[{\citenamefont{Chen et~al.}(2014)\citenamefont{Chen, Fidkowski, and
  Vishwanath}}]{Chen14}
\bibinfo{author}{\bibfnamefont{X.}~\bibnamefont{Chen}},
  \bibinfo{author}{\bibfnamefont{L.}~\bibnamefont{Fidkowski}},
  \bibnamefont{and}
  \bibinfo{author}{\bibfnamefont{A.}~\bibnamefont{Vishwanath}},
  \bibinfo{journal}{Phys. Rev. B} \textbf{\bibinfo{volume}{89}},
  \bibinfo{pages}{165132} (\bibinfo{year}{2014}).

\bibitem[{\citenamefont{Milovanovi\ifmmode~\acute{c}\else \'{c}\fi{}
  et~al.}(2016)\citenamefont{Milovanovi\ifmmode~\acute{c}\else \'{c}\fi{},
  \ifmmode \acute{C}\else \'{C}\fi{}iri\ifmmode~\acute{c}\else \'{c}\fi{}, and
  Juri\ifmmode \check{c}\else \v{c}\fi{}i\ifmmode~\acute{c}\else
  \'{c}\fi{}}}]{Milovanovic16}
\bibinfo{author}{\bibfnamefont{M.~V.}
  \bibnamefont{Milovanovi\ifmmode~\acute{c}\else \'{c}\fi{}}},
  \bibinfo{author}{\bibfnamefont{M.~D.} \bibnamefont{\ifmmode \acute{C}\else
  \'{C}\fi{}iri\ifmmode~\acute{c}\else \'{c}\fi{}}}, \bibnamefont{and}
  \bibinfo{author}{\bibfnamefont{V.}~\bibnamefont{Juri\ifmmode \check{c}\else
  \v{c}\fi{}i\ifmmode~\acute{c}\else \'{c}\fi{}}}, \bibinfo{journal}{Phys. Rev.
  B} \textbf{\bibinfo{volume}{94}}, \bibinfo{pages}{115304}
  (\bibinfo{year}{2016}).

\bibitem[{\citenamefont{Mishmash et~al.}(2018)\citenamefont{Mishmash, Mross,
  Alicea, and Motrunich}}]{Mishmash18}
\bibinfo{author}{\bibfnamefont{R.~V.} \bibnamefont{Mishmash}},
  \bibinfo{author}{\bibfnamefont{D.~F.} \bibnamefont{Mross}},
  \bibinfo{author}{\bibfnamefont{J.}~\bibnamefont{Alicea}}, \bibnamefont{and}
  \bibinfo{author}{\bibfnamefont{O.~I.} \bibnamefont{Motrunich}},
  \bibinfo{journal}{Phys. Rev. B} \textbf{\bibinfo{volume}{98}},
  \bibinfo{pages}{081107} (\bibinfo{year}{2018}).

\bibitem[{\citenamefont{Banerjee et~al.}(2018)\citenamefont{Banerjee, Heiblum,
  Umansky, Feldman, Oreg, and Stern}}]{Banerjee2018}
\bibinfo{author}{\bibfnamefont{M.}~\bibnamefont{Banerjee}},
  \bibinfo{author}{\bibfnamefont{M.}~\bibnamefont{Heiblum}},
  \bibinfo{author}{\bibfnamefont{V.}~\bibnamefont{Umansky}},
  \bibinfo{author}{\bibfnamefont{D.~E.} \bibnamefont{Feldman}},
  \bibinfo{author}{\bibfnamefont{Y.}~\bibnamefont{Oreg}}, \bibnamefont{and}
  \bibinfo{author}{\bibfnamefont{A.}~\bibnamefont{Stern}},
  \bibinfo{journal}{Nature} \textbf{\bibinfo{volume}{559}},
  \bibinfo{pages}{205} (\bibinfo{year}{2018}).

\bibitem[{\citenamefont{Balram et~al.}(2018)\citenamefont{Balram, Barkeshli,
  and Rudner}}]{Balram2018}
\bibinfo{author}{\bibfnamefont{A.~C.} \bibnamefont{Balram}},
  \bibinfo{author}{\bibfnamefont{M.}~\bibnamefont{Barkeshli}},
  \bibnamefont{and} \bibinfo{author}{\bibfnamefont{M.~S.}
  \bibnamefont{Rudner}}, \bibinfo{journal}{Phys. Rev. B}
  \textbf{\bibinfo{volume}{98}}, \bibinfo{pages}{035127}
  (\bibinfo{year}{2018}).

\bibitem[{\citenamefont{Srednicki}(1993)}]{srednicki93}
\bibinfo{author}{\bibfnamefont{M.}~\bibnamefont{Srednicki}},
  \bibinfo{journal}{Phys. Rev. Lett.} \textbf{\bibinfo{volume}{71}},
  \bibinfo{pages}{666} (\bibinfo{year}{1993}).

\bibitem[{\citenamefont{Kitaev and Preskill}(2006)}]{kitaev06b}
\bibinfo{author}{\bibfnamefont{A.~Y.} \bibnamefont{Kitaev}} \bibnamefont{and}
  \bibinfo{author}{\bibfnamefont{J.}~\bibnamefont{Preskill}},
  \bibinfo{journal}{Phys. Rev. Lett.} \textbf{\bibinfo{volume}{96}},
  \bibinfo{pages}{110404} (\bibinfo{year}{2006}).

\bibitem[{\citenamefont{Levin and Wen}(2006)}]{Levin06}
\bibinfo{author}{\bibfnamefont{M.~A.} \bibnamefont{Levin}} \bibnamefont{and}
  \bibinfo{author}{\bibfnamefont{X.-G.} \bibnamefont{Wen}},
  \bibinfo{journal}{Phys. Rev. Lett.} \textbf{\bibinfo{volume}{96}},
  \bibinfo{pages}{110405} (\bibinfo{year}{2006}).

\bibitem[{\citenamefont{Zozulya et~al.}(2007)\citenamefont{Zozulya, Haque,
  Schoutens, and Rezayi}}]{Zozulya07}
\bibinfo{author}{\bibfnamefont{O.~S.} \bibnamefont{Zozulya}},
  \bibinfo{author}{\bibfnamefont{M.}~\bibnamefont{Haque}},
  \bibinfo{author}{\bibfnamefont{K.}~\bibnamefont{Schoutens}},
  \bibnamefont{and} \bibinfo{author}{\bibfnamefont{E.~H.}
  \bibnamefont{Rezayi}}, \bibinfo{journal}{Phys. Rev. B}
  \textbf{\bibinfo{volume}{76}}, \bibinfo{pages}{125310}
  (\bibinfo{year}{2007}).

\bibitem[{\citenamefont{Li and Haldane}(2008)}]{Li08}
\bibinfo{author}{\bibfnamefont{H.}~\bibnamefont{Li}} \bibnamefont{and}
  \bibinfo{author}{\bibfnamefont{F.~D.~M.} \bibnamefont{Haldane}},
  \bibinfo{journal}{Phys. Rev. Lett.} \textbf{\bibinfo{volume}{101}},
  \bibinfo{pages}{010504} (\bibinfo{year}{2008}).

\bibitem[{\citenamefont{Chandran et~al.}(2011)\citenamefont{Chandran, Hermanns,
  Regnault, and Bernevig}}]{Chandran11}
\bibinfo{author}{\bibfnamefont{A.}~\bibnamefont{Chandran}},
  \bibinfo{author}{\bibfnamefont{M.}~\bibnamefont{Hermanns}},
  \bibinfo{author}{\bibfnamefont{N.}~\bibnamefont{Regnault}}, \bibnamefont{and}
  \bibinfo{author}{\bibfnamefont{B.~A.} \bibnamefont{Bernevig}},
  \bibinfo{journal}{Phys. Rev. B} \textbf{\bibinfo{volume}{84}},
  \bibinfo{pages}{205136} (\bibinfo{year}{2011}).

\bibitem[{\citenamefont{Qi et~al.}(2012)\citenamefont{Qi, Katsura, and
  Ludwig}}]{Qi12}
\bibinfo{author}{\bibfnamefont{X.-L.} \bibnamefont{Qi}},
  \bibinfo{author}{\bibfnamefont{H.}~\bibnamefont{Katsura}}, \bibnamefont{and}
  \bibinfo{author}{\bibfnamefont{A.~W.~W.} \bibnamefont{Ludwig}},
  \bibinfo{journal}{Phys. Rev. Lett.} \textbf{\bibinfo{volume}{108}},
  \bibinfo{pages}{196402} (\bibinfo{year}{2012}).

\bibitem[{\citenamefont{Zhao et~al.}(2011)\citenamefont{Zhao, Sheng, and
  Haldane}}]{Zhao11}
\bibinfo{author}{\bibfnamefont{J.}~\bibnamefont{Zhao}},
  \bibinfo{author}{\bibfnamefont{D.~N.} \bibnamefont{Sheng}}, \bibnamefont{and}
  \bibinfo{author}{\bibfnamefont{F.~D.~M.} \bibnamefont{Haldane}},
  \bibinfo{journal}{Phys. Rev. B} \textbf{\bibinfo{volume}{83}},
  \bibinfo{pages}{195135} (\bibinfo{year}{2011}).

\bibitem[{\citenamefont{Biddle et~al.}(2011)\citenamefont{Biddle, Peterson, and
  Das~Sarma}}]{Biddle11}
\bibinfo{author}{\bibfnamefont{J.}~\bibnamefont{Biddle}},
  \bibinfo{author}{\bibfnamefont{M.~R.} \bibnamefont{Peterson}},
  \bibnamefont{and}
  \bibinfo{author}{\bibfnamefont{S.}~\bibnamefont{Das~Sarma}},
  \bibinfo{journal}{Phys. Rev. B} \textbf{\bibinfo{volume}{84}},
  \bibinfo{pages}{125141} (\bibinfo{year}{2011}).

\bibitem[{\citenamefont{Lewenstein et~al.}(2007)\citenamefont{Lewenstein,
  Sanpera, Ahufinger, Damski, Sen, and Sen}}]{Lewenstein2007}
\bibinfo{author}{\bibfnamefont{M.}~\bibnamefont{Lewenstein}},
  \bibinfo{author}{\bibfnamefont{A.}~\bibnamefont{Sanpera}},
  \bibinfo{author}{\bibfnamefont{V.}~\bibnamefont{Ahufinger}},
  \bibinfo{author}{\bibfnamefont{B.}~\bibnamefont{Damski}},
  \bibinfo{author}{\bibfnamefont{A.}~\bibnamefont{Sen}}, \bibnamefont{and}
  \bibinfo{author}{\bibfnamefont{U.}~\bibnamefont{Sen}}, \bibinfo{journal}{Adv.
  Phys.} \textbf{\bibinfo{volume}{56}}, \bibinfo{pages}{243}
  (\bibinfo{year}{2007}).

\bibitem[{\citenamefont{Bloch et~al.}(2008)\citenamefont{Bloch, Dalibard, and
  Zwerger}}]{Bloch2008}
\bibinfo{author}{\bibfnamefont{I.}~\bibnamefont{Bloch}},
  \bibinfo{author}{\bibfnamefont{J.}~\bibnamefont{Dalibard}}, \bibnamefont{and}
  \bibinfo{author}{\bibfnamefont{W.}~\bibnamefont{Zwerger}},
  \bibinfo{journal}{Rev. Mod. Phys.} \textbf{\bibinfo{volume}{80}},
  \bibinfo{pages}{885} (\bibinfo{year}{2008}).

\bibitem[{\citenamefont{Dalibard et~al.}(2011)\citenamefont{Dalibard, Gerbier,
  Juzeli\ifmmode~\bar{u}\else \={u}\fi{}nas, and \"Ohberg}}]{Dalibard2011}
\bibinfo{author}{\bibfnamefont{J.}~\bibnamefont{Dalibard}},
  \bibinfo{author}{\bibfnamefont{F.}~\bibnamefont{Gerbier}},
  \bibinfo{author}{\bibfnamefont{G.}~\bibnamefont{Juzeli\ifmmode~\bar{u}\else
  \={u}\fi{}nas}}, \bibnamefont{and}
  \bibinfo{author}{\bibfnamefont{P.}~\bibnamefont{\"Ohberg}},
  \bibinfo{journal}{Rev. Mod. Phys.} \textbf{\bibinfo{volume}{83}},
  \bibinfo{pages}{1523} (\bibinfo{year}{2011}).

\bibitem[{\citenamefont{Eckardt}(2017)}]{Andre2017}
\bibinfo{author}{\bibfnamefont{A.}~\bibnamefont{Eckardt}},
  \bibinfo{journal}{Rev. Mod. Phys.} \textbf{\bibinfo{volume}{89}},
  \bibinfo{pages}{011004} (\bibinfo{year}{2017}).

\bibitem[{\citenamefont{Fano et~al.}(1986)\citenamefont{Fano, Ortolani, and
  Colombo}}]{Fano86}
\bibinfo{author}{\bibfnamefont{G.}~\bibnamefont{Fano}},
  \bibinfo{author}{\bibfnamefont{F.}~\bibnamefont{Ortolani}}, \bibnamefont{and}
  \bibinfo{author}{\bibfnamefont{E.}~\bibnamefont{Colombo}},
  \bibinfo{journal}{Phys. Rev. B} \textbf{\bibinfo{volume}{34}},
  \bibinfo{pages}{2670} (\bibinfo{year}{1986}).

\bibitem[{\citenamefont{Sreejith et~al.}(2017)\citenamefont{Sreejith, Zhang,
  and Jain}}]{Sreejith17}
\bibinfo{author}{\bibfnamefont{G.~J.} \bibnamefont{Sreejith}},
  \bibinfo{author}{\bibfnamefont{Y.}~\bibnamefont{Zhang}}, \bibnamefont{and}
  \bibinfo{author}{\bibfnamefont{J.~K.} \bibnamefont{Jain}},
  \bibinfo{journal}{Phys. Rev. B} \textbf{\bibinfo{volume}{96}},
  \bibinfo{pages}{125149} (\bibinfo{year}{2017}).

\bibitem[{\citenamefont{Dimov et~al.}(2008)\citenamefont{Dimov, Halperin, and
  Nayak}}]{Dimov08}
\bibinfo{author}{\bibfnamefont{I.}~\bibnamefont{Dimov}},
  \bibinfo{author}{\bibfnamefont{B.~I.} \bibnamefont{Halperin}},
  \bibnamefont{and} \bibinfo{author}{\bibfnamefont{C.}~\bibnamefont{Nayak}},
  \bibinfo{journal}{Phys. Rev. Lett.} \textbf{\bibinfo{volume}{100}},
  \bibinfo{pages}{126804} (\bibinfo{year}{2008}).

\bibitem[{\citenamefont{Biddle et~al.}(2013)\citenamefont{Biddle, Peterson, and
  Das~Sarma}}]{Biddle13}
\bibinfo{author}{\bibfnamefont{J.}~\bibnamefont{Biddle}},
  \bibinfo{author}{\bibfnamefont{M.~R.} \bibnamefont{Peterson}},
  \bibnamefont{and}
  \bibinfo{author}{\bibfnamefont{S.}~\bibnamefont{Das~Sarma}},
  \bibinfo{journal}{Phys. Rev. B} \textbf{\bibinfo{volume}{87}},
  \bibinfo{pages}{235134} (\bibinfo{year}{2013}).

\bibitem[{\citenamefont{{Stern} et~al.}(2010)\citenamefont{{Stern},
  {Plochocka}, {Umansky}, {Maude}, {Potemski}, and {Bar-Joseph}}}]{Stern10}
\bibinfo{author}{\bibfnamefont{M.}~\bibnamefont{{Stern}}},
  \bibinfo{author}{\bibfnamefont{P.}~\bibnamefont{{Plochocka}}},
  \bibinfo{author}{\bibfnamefont{V.}~\bibnamefont{{Umansky}}},
  \bibinfo{author}{\bibfnamefont{D.~K.} \bibnamefont{{Maude}}},
  \bibinfo{author}{\bibfnamefont{M.}~\bibnamefont{{Potemski}}},
  \bibnamefont{and}
  \bibinfo{author}{\bibfnamefont{I.}~\bibnamefont{{Bar-Joseph}}},
  \bibinfo{journal}{Phys. Rev. Lett.} \textbf{\bibinfo{volume}{105}},
  \bibinfo{pages}{096801} (\bibinfo{year}{2010}).

\bibitem[{\citenamefont{Rhone et~al.}(2011)\citenamefont{Rhone, Yan, Gallais,
  Pinczuk, Pfeiffer, and West}}]{Rhone10}
\bibinfo{author}{\bibfnamefont{T.~D.} \bibnamefont{Rhone}},
  \bibinfo{author}{\bibfnamefont{J.}~\bibnamefont{Yan}},
  \bibinfo{author}{\bibfnamefont{Y.}~\bibnamefont{Gallais}},
  \bibinfo{author}{\bibfnamefont{A.}~\bibnamefont{Pinczuk}},
  \bibinfo{author}{\bibfnamefont{L.}~\bibnamefont{Pfeiffer}}, \bibnamefont{and}
  \bibinfo{author}{\bibfnamefont{K.}~\bibnamefont{West}},
  \bibinfo{journal}{Phys. Rev. Lett.} \textbf{\bibinfo{volume}{106}},
  \bibinfo{pages}{196805} (\bibinfo{year}{2011}).

\bibitem[{\citenamefont{Tiemann et~al.}(2012)\citenamefont{Tiemann, Gamez,
  Kumada, and Muraki}}]{Tiemann828}
\bibinfo{author}{\bibfnamefont{L.}~\bibnamefont{Tiemann}},
  \bibinfo{author}{\bibfnamefont{G.}~\bibnamefont{Gamez}},
  \bibinfo{author}{\bibfnamefont{N.}~\bibnamefont{Kumada}}, \bibnamefont{and}
  \bibinfo{author}{\bibfnamefont{K.}~\bibnamefont{Muraki}},
  \bibinfo{journal}{Science} \textbf{\bibinfo{volume}{335}},
  \bibinfo{pages}{828} (\bibinfo{year}{2012}).

\end{thebibliography}
\end{document}